\documentclass[12pt,a4paper]{article}

\usepackage[a4paper]{geometry}
\usepackage{graphicx}
\usepackage[textwidth=8em,textsize=small]{todonotes}
\usepackage{amsmath}
\usepackage{natbib}
\usepackage{subfigure}
\usepackage{amsfonts}
\usepackage{amssymb}
\usepackage{mathtools}
\usepackage{multirow}
\usepackage{appendix}
\usepackage{indentfirst}
\usepackage[english]{babel}
\usepackage{hyperref}
\hypersetup{colorlinks=true,
            linkcolor=blue,
            filecolor=blue,      
            urlcolor=blue,
            citecolor=blue}
            
\newcommand{\bfvarepsilon}{\mbox{\boldmath $\varepsilon$}}

\newcommand{\bftheta}{\mbox{\boldmath $\theta$}}
\newcommand{\bfbeta}{\mbox{\boldmath $\beta$}}
\newcommand{\bfTheta}{\mbox{\boldmath $\Theta$}}

\newcommand{\bfomega}{\mbox{\boldmath $\omega$}}

\providecommand{\keywords}[1]{\textbf{\textbf{Keywords:}} #1}
\newcommand{\bfs}{\mathbf{s}}
\newcommand{\mathbbm}[1]{\text{\usefont{U}{bbm}{m}{n}#1}}
\title{Space-time calibration of wind speed forecasts from regional climate models}

\author{Luiz E. S. Gomes$^1$ \and Tha\'is C. O. Fonseca$^{1,2}$\footnote{\textit{Address for correspondence}: Tha\'is C. O. Fonseca, Department of Statistics,
University of Warwick, Coventry, CV4 7AL, United Kingdom. E-mail: Thais.Oliveira-Da-Fonseca@warwick.ac.uk} \and Kelly C. M. Gon\c{c}alves$^1$ \and Ramiro Ruiz-C\'ardenas$^3$}

\date{
    $^1${Federal University of Rio de Janeiro, Brazil}\\
    $^2${University of Warwick, United Kingdom}\\
    $^3${Consultant, Brazil}
}

\begin{document}

\maketitle

\begin{abstract}
Numerical weather predictions (NWPs) are systematically subject to errors due to the deterministic solutions used by numerical models to simulate the atmosphere. Statistical postprocessing techniques are widely used nowadays for NWP calibration. However, time-varying bias is usually not accommodated by such models. The calibration performance is also sensitive to the temporal window used for training. This paper proposes space-time models that extend the main statistical postprocessing approaches to calibrate NWP model outputs. Trans-Gaussian random fields are considered to account for meteorological variables with asymmetric behavior. Data augmentation is used to account for censoring of the response variable. The benefits of the proposed extensions are illustrated through the calibration of hourly 10-meter height wind speed forecasts in Southeastern Brazil coming from the Eta model.

\end{abstract}

% --- Section 1
\section{Introduction}
\label{sec:intro}

Numerical weather predictions (NWPs) are often based on mathematical models that make deterministic predictions from current atmospheric conditions. Such models rely on complex systems of differential equations that do not have an analytical solution. Therefore, numerical integration is used to simulate the physical, dynamic, and thermodynamic processes of the atmosphere depending on its current state. From this initial solution, it is possible to solve the system for any future time of interest \citep{krishnamurti1995numerical}.

These numerical systems are solved in a discrete grid which means that the model assumes uniform predictions for any point in the region belonging to a given cell of the grid. In particular, forecasts at each grid cell are based on average data in its region (e.g., average elevation and predominant vegetation). Under these assumptions, the representativeness of predictions in cells with, for example, complex orography, dense vegetation, or presence of water bodies, is inadequate due to differences between the actual characteristics of the surface and the homogenization made by the model. Therefore, outputs from NWP models may not be representative at specific locations \citep{chou2007refinamento}, thus producing systematic errors.

Meteorological phenomena are often not well described by a single numerical prediction. As an alternative, ensembles of forecasts, i.e., groups of outputs coming from multiple runs of either, the same or several models under varying initial conditions and model physics, are considered to allow quantification of uncertainty. The ensembles can be interpreted as a Monte Carlo experiment aiming to produce a range of future states of the atmosphere from different initial conditions \citep{epstein1969stochastic}. Moreover, the simulation of scenarios can indicate extreme events that would not be identified by just one run of the numerical model \citep{grimit2007measuring}. However, according to \cite{gneiting2005calibrated}, ensemble forecasts are often under-dispersive, i.e., the ensemble spread is too narrow to account for all the uncertainty. As a solution, calibration is often used to correct these error patterns.

Statistical postprocessing techniques have been widely used in recent years to minimize the above limitations of numerical models and to enhance both, the reliability and statistical consistency of numerical weather predictions. Many methods for the statistical calibration of ensemble forecasts are now available. Recent reviews on this subject can be found in \cite{le2017review} and \cite{vannitsem2018postprocessing}. The pioneer postprocessing technique in this context was the application of model output statistics \citep[MOS,][]{glahn1972use} which considers a multiple linear regression relating the responses (e.g., observed wind) to the set of ensemble members, assuming constant variance. Several extensions of this precursor method have been proposed, such as updatable MOS \citep[UMOS,][]{wilson2002canadian} which varies the size of the training period; MOC \citep{mao1999optimal} which directly models the prediction error; and the generalized MOS \citep[e.g.,][]{piani2010statistical} which considers generalized linear models instead of Gaussian distributions for the response vector. An extension of the MOS technique known as ensemble MOS \citep[EMOS,][]{gneiting2005calibrated} allows for a spread-skill relationship between the dispersion of ensemble members and the response variance \citep{whitaker1998relationship}. In the context of spatial calibration, the geostatistical output perturbation \citep[GOP,][]{gel2004calibrated} allows spatial dependence, resulting in calibrated weather fields for a fixed temporal horizon. Moreover, the spatial EMOS method \citep[SEMOS,][]{feldmann2015spatial} combines the EMOS and GOP techniques.

Estimation of parameters in statistical postprocessing usually occurs in a subset of the data, called a training period, defined through a moving time window that accounts for the effect of both past observations and numerical predictions. If the training period is reasonably long, it becomes easier to estimate the uncertainty of predictions \citep{gneiting2014calibration}. However, longer training periods may introduce distortions in the calibration due to seasonal effects. This challenging tradeoff suggests that the window size must be tailored for the specific application to achieve good results. To illustrate, \cite{raftery2005using} analyzed the effects of the training window size on the uncertainty estimation of parameters in the calibration of temperature and sea level pressure predictions coming from the same numerical model. The authors reported gains with the use of a time window as large as 25 days for both case studies and highlighted the need for an automatic way to choose the length of the training period.

The above shortcomings of current statistical postprocessing approaches can be addressed by the incorporation of statistical techniques such as spatially structured calibration models. This approach may be useful in correcting NWP errors in regions where numerical models have smoothed important characteristics of the terrain, while the Kalman filter \citep{kalman1960new} and Bayesian dynamic models \citep{west1997bayesian} are natural alternatives to account for seasonality and temporal dynamics of bias in postprocessing models.

The proposed models should also be able to accommodate the spatial characteristics of the terrain, as well as fairly skew and censor the behavior of the response, which can be essential for prediction and uncertainty quantification of certain variables, such as wind speed. In this context, the transformed Gaussian dynamic model has been useful in applied settings, such as precipitation modeling \citep{bardossy1992space, sanso1999venezuelan}.  \cite{sigrist2012dynamic} and \cite{sigrist2015stochastic} also followed the latent Gaussian approach of \cite{bardossy1992space} to predict weather variables. In particular, \cite{sigrist2012dynamic} proposed a spatiotemporal model for rainfall data based on the convolution of covariance functions. Note that in this case, the resulting covariance structure is usually not obtained in closed form. \cite{sigrist2015stochastic} proposed a stochastic partial differential equation-based model and applied it to postprocessing of precipitation forecasts. The model is based on the spectral domain. For further details about the spectral approach to spatiotemporal modeling, see \cite{Fuentes08}.

In general, the proposed model is a known spatial dynamic regression model, which we have extended to include Box-Cox transformation to adapt to the wind speed characteristics such as asymmetry and censoring. Developing asymmetric models for spatiotemporal applications is non-trivial, as discussed in recent papers. \cite{genton2012identifiability} mentioned that the parameters in asymmetric spatial processes are not well identified, even if the number of locations is large. \cite{zhang2010spatial} indicated that for skew-normal spatial models when the asymmetry is high, the spatial correlation between two locations approaches 1 regardless of the distance between these two locations. In this context, the proposed model is an interesting alternative that accounts for non-stationarity in time, asymmetry, censoring, and spatiotemporal correlation. The Gaussian model and conditional Gaussian extensions are usual choices for the spatiotemporal setup due to the validity of the process and analytical results available for fast predictions. In the context of spatial generalized linear models \citep{diggle2007springer}, the model would be hierarchical, and the spatiotemporal random effects would be assumed in the second level of this hierarchy. We did not follow this approach, as it would be computationally even more demanding than the conditional Gaussian alternative with Box-Cox transformation.

This paper proposes a unifying approach to statistical postprocessing by accounting for both spatial and temporal dependencies overcoming the need for large time windows in the definition of training sets. Our proposal is inspired by current postprocessing approaches, which are extended based on both, spatial dynamic linear models and data augmentation techniques. An optimized Markov chain Monte Carlo (MCMC) scheme based on the robust adaptive Metropolis algorithm \citep{vihola2012robust} is used to perform efficient statistical inference and prediction.

The primary motivation of this study is the calibration of 10-meter height wind speed forecasts since that meteorological variable presents some particular features that might not be properly addressed by the usual postprocessing techniques. According to \cite{ailliot2006autoregressive}, some of these features include: intermittent atmospheric regimes with a predominance of a certain wind direction in given regions; spatial and temporal correlation; non-Gaussianity; non-stationarity; conditional heteroscedasticity, i.e., the variance of the wind speed changes frequently in time; seasonal annual and diurnal components due to the effects of the sun and seasons; and possible trends.

The remainder of this article is organized as follows. Section \ref{sec:data} presents the wind speed data which motivates our proposed modeling approach. Section \ref{sec:postmodels} briefly describes the main postprocessing approaches used to calibrate numerical weather forecasts; Section \ref{subsec:proposed} presents the proposed extensions to enhance current postprocessing techniques and describes the inference procedure. The performance of the proposed extensions for applications using real data and simulated experiments is evaluated and compared with the current approaches in Sections \ref{sec:application} and \ref{sec:applicationsimul}, respectively. The article concludes with a discussion in Section \ref{sec:discussion}.

% --- Section 2
\section{Wind speed data for the state of Minas Gerais}
\label{sec:data}

The main motivation for the proposed models presented in this paper is the calibration of wind speed forecasts generated by the Eta regional climate model for the state of Minas Gerais, Brazil. The state, located in the Southeast region of Brazil, is formed by abundant plateaus interspersed with rugged relief, ranging from 100 to 2800 meters above sea level, which provides to the state exceptional water resources. The predominant vegetation, known as \textit{Cerrado}(savanna), is characterized by large variations in the landscape between the rainy and dry seasons. This results in a seasonal influence of the surface roughness on the displacement of the winds, which are stronger during winter and spring. The climate in Minas Gerais varies from hot semiarid to humid mesothermal. Rainfall distribution is uneven, with the northern region having long periods of drought and the highest temperatures, while the southern region (high elevation areas) concentrates the highest total annual rainfall \citep{amarante2010atlas}.

Two sources of hourly 10-meter height wind speed data for the region of interest, covering the two years from 1 October 2015 to 30 September 2017, were used in this application: ground measurements and numerical forecasts. The measurement data are recorded by an irregular network of weather stations operated by the Brazilian National Institute of Meteorology (INMET), available at $<$ \url{http://www.inmet.gov.br} $>$. A set of 68 stations from this network, spread over Minas Gerais and its surroundings were selected with the requirement of at least 70\% of data availability. In these data, distances in space range from 7.9 to 1145 km, and elevation of stations varies from 60 to 1359 m. The numerical accuracy of the data is to one decimal place.

\begin{figure}
 \centering
 \makebox{\includegraphics[scale=.4,angle=-90]{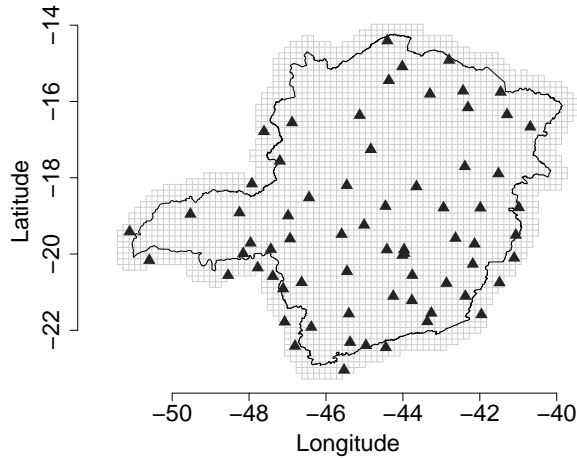}}
 \caption{Map of Minas Gerais showing the location of the 68 weather stations (solid triangles). The solid lines represent the regular grid of the Eta regional climate model with a resolution of 15 km.}
 \label{fig:locations}
\end{figure}

Numerical forecasts are generated by the Eta model \citep{mesinger1988step,black1994new}. It is a regional climate model run at the Center for Weather Forecasting and Climate Studies (CPTEC). Its extent covers most of South America and Central America. The model results are disclosed twice a day at 0000 UTC and 1200 UTC. We used hourly Eta model outputs from the run began at 1200 UTC with a horizontal resolution of 15 km, 50 vertical layers, and a lead time of up to 264 hours (11 days). This lead time allows obtaining up to 10 different wind speed forecasts for each hour, coming from the daily runs of the model. These groups of forecasts composed our ensembles in this application. This was because just one version of the model is issued at each running time. Hence, ensembles of forecasts for several numerical models were not available. Figure \ref{fig:locations} presents the distribution of selected stations in the region of interest as well as the discrete grid of the numerical model.

Gridded forecasts were bilinearly interpolated to the locations of the 68 stations, to form the dataset with hourly 10-meter height wind speed forecasts for calibration at these locations. According to \cite{gel2004calibrated}, more complex interpolation methods can be used but these are unlikely to produce considerable gains if the grid is very fine.

\begin{figure}
 \centering
 \makebox{\includegraphics[scale=.5, angle=-90]{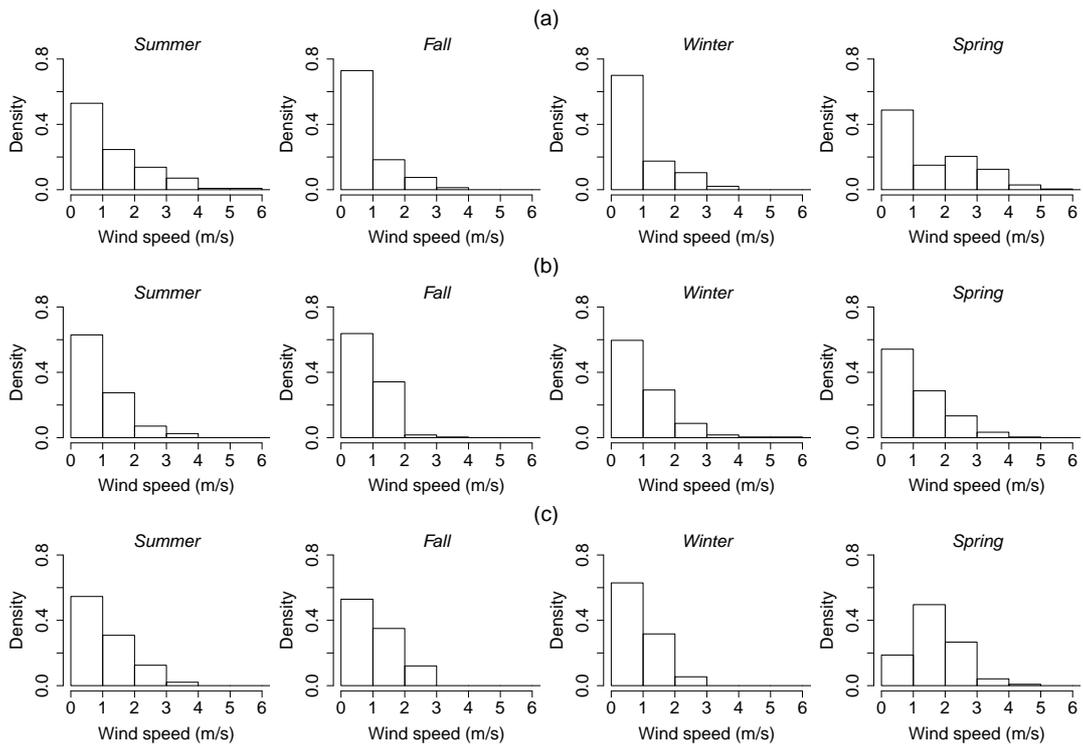}}
  \caption{Histograms of observed wind speed at 10-meter height during the seasons from 1 October 2015 to 30 September 2017 at (a) Vi\c cosa, (b) Muria\'e, and (c) Pampulha (Belo Horizonte) weather stations.}
  \label{fig:data}
\end{figure}

To illustrate the local characteristics of wind speed, Figure \ref{fig:data} presents some histograms with the wind speed distribution during the seasons at three selected stations: Vi\c cosa, Muria\'e, and Pampulha (Belo Horizonte). The histograms clearly show an asymmetric distribution with a frequent point mass at zero. 
In general, lower wind speeds are recorded during the fall and summer and this pattern extends to the vast majority of available weather station readings in the dataset. Particularly, the wind speed recorded during the spring has the highest threshold and the smallest point mass at zero. This wind regime is even more evident at the Pampulha (Belo Horizonte) station, as illustrated in Figure \ref{fig:data}(c). These specific aspects of wind speed distribution motivate the space-time postprocessing models proposed in this paper, as well as the need to consider the left censoring and data transformation.

% A510 - Vi\c cosa
% A517 - Muria\'e
% A521 - Pampulha, Belo Horizonte

% --- Section 3
\section{Statistical postprocessing models}
\label{sec:postmodels}

We now describe the main postprocessing models used in the probabilistic calibration of forecasts generated by NWP models. Our proposed methods based on dynamic spatial models are presented as an automatic bias correction alternative for ensemble forecasting in longer training periods.

\subsection{Static calibration models}

The model output statistics \citep[MOS,][]{glahn1972use} defines a relation between
an ensemble with $m$ forecasts $F_1,...,F_m$, individually distinguishable, for an univariate quantity of interest $Y$ through a multiple linear regression model given by:
\begin{equation} 
\label{eq:MOS}
Y = \theta_0 + \theta_1F_1 + ... + \theta_mF_m + \varepsilon,
\end{equation}
where the error term $\varepsilon$ assures that $E(\varepsilon)=0$ and  $Var(\varepsilon) = \sigma^2$. 

An extension of MOS called ensemble MOS \citep[EMOS,][]{gneiting2005calibrated}, also known as non-homogeneous regression, allows the response variance to depend on the dispersion of ensemble members. This improvement is called the spread-skill relationship. It is based on the premise of a positive relationship between ensemble spread and forecast absolute error.
The EMOS method assumes that $\varepsilon$ is such that:
\begin{equation} 
\label{eq:EMOSvar}
E(\varepsilon) = 0, \quad Var(\varepsilon) = \sigma^2_* = \beta_0 + \beta_1S^2,
\end{equation}
where $S^2$ is the sample variance of ensemble members and $\bfbeta = (\beta_0,\beta_1)'$ are non-negative coefficients. \cite{gneiting2005calibrated} assumed Gaussianity and estimated model parameters by minimization of continuous ranked probability score \citep[CRPS,][]{matheson1976scoring}. Other applications of this method in the context of wind speed, strong winds and wind direction can be found in  \cite{thorarinsdottir2010probabilistic}, \cite{thorarinsdottir2012probabilistic}, and \cite{schuhen2012ensemble}, respectively.

A spatial extension of MOS called geostatistical output perturbation \citep[GOP,][]{gel2004calibrated} is the pioneer statistical postprocessing model that considers the correlation between the measurements of a meteorological variable at distinct locations. This technique produces calibrated forecasts for entire weather fields. Let $\left\lbrace Y(\bfs), \bfs \in S \subset \mathbb{R}^2 \right\rbrace$ be a random weather field and $\textbf{Y} = \left(y(\bfs_1),...,y(\bfs_n)\right)'$, an observed sample at $n$ locations. Considering a set of $m$ ensemble members for these locations, represented by $\textbf{F}_{1} = \left(F_1(\bfs_1),...,F_1(\bfs_n)\right)', \dots , \textbf{F}_{m} = \left(F_m(\bfs_1),...,F_m(\bfs_n)\right)' $, and assuming $\bfvarepsilon = \left( \varepsilon(\bfs_1),...,\varepsilon(\bfs_n)\right)'$, a vector of observations from a Gaussian process $\left\lbrace \varepsilon(\bfs), \bfs \in S \right\rbrace$ is obtained, such that:
\begin{equation} 
\label{eq:GOPvar}
E(\bfvarepsilon) = \textbf{0}_n, \mbox{ and } \quad Cov(\varepsilon(\bfs_i),\varepsilon(\bfs_j)) = \Sigma_{i,j} = \sigma^2C(\bfs_i,\bfs_j), \quad i,j = 1,\dots,n,
\end{equation}
\noindent
where $\textbf{0}_n$ is a zero $n$-vetor and $C(\cdot,\cdot)$ is a valid spatial correlation function \citep{cressie1993statistics} for any pair of locations in $S$. This method is a multivariate extension of the model described in (\ref{eq:MOS}) and therefore also does not allocate the ensemble information in its structure.

Finally, considering simultaneously the ensemble information and the spatial structure of meteorological variables, the spatial EMOS \citep[SEMOS,][]{feldmann2015spatial} is presented combining the EMOS and GOP models. It consists of the GOP model formulation but assumes that:
\begin{equation} 
\label{eq:SEMOSvar}
Cov(\varepsilon(\bfs_i),\varepsilon(\bfs_j)) = \Sigma^*_{i,j} = D_{i,i}C(s_i,s_j)D_{j,j}, \quad i,j = 1,...,n,
\end{equation}
\noindent
where $D = \text{diag}(\sqrt{\beta_0 + \beta_1 S^2_1},...,\sqrt{\beta_0 + \beta_1 S^2_n})$ is an $n\times n$ diagonal matrix with $S^2_i$ representing the sample variance of ensemble members for location $i$. The models presented here, in their canonical form, are not appropriate for calibration of asymmetric variables (e.g., wind speed and rainfall).

\subsection{Proposed dynamic calibration models}
\label{subsec:proposed}

The postprocessing models proposed here improve previous approaches in several ways, accounting for both, temporal dynamics and asymmetric distributions with mass at zero through a spatiotemporal modeling approach with censoring. These novel methods are presented next. In particular, the truncated Gaussian model which has been used in precipitation prediction \citep{bardossy1992space, sanso1999venezuelan} is presented in the context of forecast calibration from NWP models.

\subsubsection{Spatiotemporal trans-Gaussian model}

Let $\left\lbrace Y_t(\bfs), \bfs \in S \subset \mathbb{R}^2, t=1,...,T \right\rbrace $ be a random spatial field in discrete time $t$. The observed response vector at $n$ locations $\textbf{Y}_{t} = (Y_t(\bfs_1),...,Y_t(\bfs_n))'$ is composed by censored random variables, $Y_t(\bfs_i) \geq c$, $i=1,\dots,n$, $t=1,\dots,T$. Assume that $\textbf{Y}_{t}$ follows a truncated Gaussian distribution as follows:

\begin{equation}
\label{eq:BC}
Y_t(\bfs)=\left\{\begin{array}{rc}
BC^{-1} \left( X_t(\bfs), \lambda \right) ,&\mbox{if}\quad {BC^{-1} \left( X_t(\bfs), \lambda \right)} > c,\\
c, &\mbox{if}\quad BC^{-1} \left( X_t(\bfs), \lambda \right) \leq c,
\end{array}\right.
\end{equation}
where $c$ is a known constant, $\lambda$ is the unknown transformation parameter, $X_t(\bfs)$ is a Gaussian process and $BC(\cdot,\lambda)$ represents the family of Box-Cox transformations \citep{box1964analysis} defined as:
$$
BC(y,\lambda)=\left\{\begin{array}{rc}
\left( y^{\lambda}-1 \right) /\lambda ,&\mbox{if}\quad \lambda \neq 0 \quad \mbox{and} \quad y > 0,\\
\log y, &\mbox{if}\quad \lambda = 0 \quad \mbox{and} \quad y > 0. \\
\end{array}\right.
$$

Therefore, $X_t(\bfs)$ is a latent Gaussian field which allows for asymmetry in the resulting process of interest $Y_t(\bfs)$. That is, after transforming the possibly asymmetric process $Y_t(\bfs)$, the resulting field will follow a Gaussian process. The trans-Gaussian model considered adds flexibility to the Gaussian model usually assumed for random spatiotemporal fields. This strategy was considered due to its simple specification as opposed to other alternatives to develop non-Gaussian processes which may result in invalid processes \cite[see for instance][]{zhang2010spatial}. The proposed trans-Gaussian field accounts for asymmetry (e.g., the log transformation) which is a known characteristic of wind speed data modeling.

\begin{figure}
 \centering
 \makebox{\includegraphics[scale=.5, angle=-90]{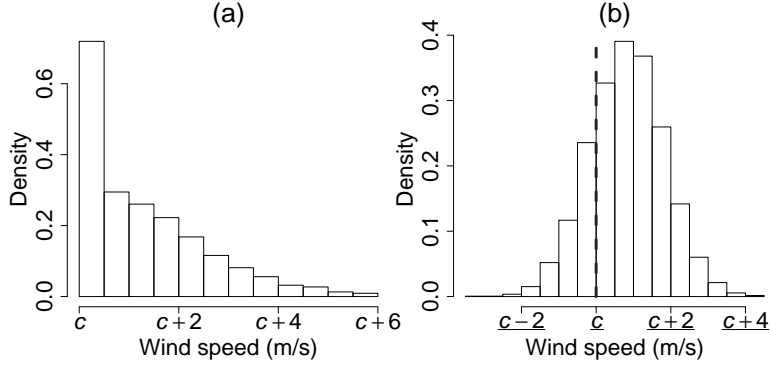}}
 \caption{Histograms for (a) simulated wind speed and (b) transformed simulated wind speed by Box-Cox transformation as defined in (\ref{eq:BC}). The dashed line represents the transformed threshold value \underline{$c$} $= BC(c, \lambda)$.}
 \label{fig:censoring}
\end{figure}

Furthermore, censoring is admitted through a known constant $c$. In the context of wind speed calibration, $c$ represents the minimum allowed value for observed wind speed (usually 0 m/s). However, depending on the application context, this value can be greater than zero. Figure \ref{fig:censoring} illustrates the practical effect of this modeling setup on simulated wind speed values at a single location. Note that values below \underline{$c$} from transformed and truncated Gaussian latent process induce a mass at $c$ in the process of interest $Y_t(\bfs)$. Application of the complete model with Box-Cox transformation and censoring is facilitated by the use of data augmentation techniques \citep{tanner1987calculation}. It also naturally allows for missing data. 

Spatiotemporal trans-Gaussian models with different transformation functions are widely used, especially in the context of precipitation forecasting. Specific transformation functions are defined for each application. In general, the power transform function \citep[e.g.,][]{sanso2004bayesian, ailliot2009space} and a quadratic form of the power transform function \citep[e.g.,][]{allcroft2003latent} are the most used due to their simplicity. However, other more complex functions can be found, such as the power-exponential transformation \citep[e.g.,][]{allard2015disaggregating}. For the wind speed prediction context, we selected the Box-Cox transformation, a power transform function extension, often used in temporal and spatial applications as discussed, respectively, in \cite{guerrero1993time} and \cite{de1997bayesian}.

In the usual setup for statistical calibration of numerical models time windows are defined to account for observed and predicted temporal variation, the so-called training period. Although larger training windows result in smaller uncertainty, they introduce distortions due to seasonal effects. In general, the seasonal patterns of meteorological variables are well defined (e.g., solar forcing -- 24 hours and seasons -- 3 months). Thus, the Kalman filter \citep{kalman1960new} allows for both longer temporal training periods and the inclusion of temporal dynamics in the bias parameters. In the following, we consider the Bayesian dynamic approach which considers the Kalman filter for forward filtering and backward posterior smoothing for a fully Bayesian inference procedure as proposed by \cite{west1997bayesian}. This approach allows for time-varying regression coefficients while accounting for spatial dependence in the region of interest, overcoming the over-smoothing of numerical forecasts.

\subsubsection{Dynamic geostatistical output perturbation}

The proposed model adds a temporal dynamic to the GOP formulation by allowing the coefficients to vary over time. The covariance modeling is stochastic over time with Beta-Gamma evolution depending on discount factors. Assuming the structure stated in (\ref{eq:BC}) for the transformation of the original process $Y_t(s)$, the dynamics are introduced by defining the observation and state equations as follows:
\label{eq:DGOP}
\begin{align}
\textbf{X}_{t}= {\textbf{F}'_{t}}\bftheta_t + \bfvarepsilon_t, \quad & \bfvarepsilon_t \sim N(\textbf{0}_n,\Sigma_t), \quad \Sigma_t = \varphi_t^{-1}H, \label{eq:DGOPxt}\\
\bftheta_t= \textbf{G}_t\bftheta_{t-1} + \bfomega_t, \quad & \bfomega_t \sim T_{n_{t-1}}(\textbf{0}_p,\textbf{W}_t),\label{eq:DGOPthetat}\\
\varphi_t = \gamma_t \varphi_{t-1}/\delta_V, \quad & \gamma_t \sim \text{Beta}(\kappa_t,\bar{\kappa}_t),\label{eq:DGOPpsit}
% \kappa_t = \delta_V n_{t-1}/2 \mbox{ and } & \bar{\kappa}_t = (1-\delta_V) n_{t-1}/2
\end{align}
where $\bfvarepsilon_t = \left( \varepsilon_t(\bfs_1),...,\varepsilon_t(\bfs_n)\right)'$ follows a zero mean multivariate normal distribution with covariance matrix $\Sigma_t = \varphi_t^{-1}H$ such that $\varphi_t=1/\sigma^2_t$ is the precision and $H$ is the correlation matrix with elements $H_{{i,j}} = C(\bfs_i,\bfs_j),\quad i,j = 1,\dots,n$. $C(\cdot,\cdot;\phi)$ is a valid correlation function depending on an unknown parameter $\phi$. For instance, a simple correlation function is $C(\bfs_i,\bfs_j;\phi) = \exp(-\phi\|{\bfs_i - \bfs_j}\|)$, with $\phi > 0$ representing the spatial
decay parameter and $\|{\bfs_i - \bfs_j}\|$, the Euclidean distance between locations $\bfs_i$ and $\bfs_j$, $i,j=1,...,n$. Other possible choices are the Mat\'ern class \citep{Matern86} and the Cauchy class \citep{Gneit04}. $\textbf{X}_{t} = \left( X_t(\bfs_1),...,X_t(\bfs_n) \right)'$, ${\textbf{F}'_{t}}$ is a matrix with dimension $n \times r$ $(r \geq m)$ composed by covariates (e.g., predicted ensembles, latitude, longitude, and elevation) and $\bftheta_t$ represents the state-space vector of variables with dimension $r$.

For the purely temporal components in (\ref{eq:DGOPthetat}) and (\ref{eq:DGOPpsit}), $\textbf{G}_t$ is an evolution matrix with dimension $r$, $\bfomega_t$ are mutually independent and follow a zero mean multivariate Student-t distribution with $n_{t-1}$ degrees of freedom and unknown scale matrix $\textbf{W}_t$ which can be estimated using discounting factors. The degrees of freedom $n_{t-1}$ and the shape parameters $\kappa_t$ and $\bar{\kappa}_t$ are defined through a Beta-Gamma stochastic evolution \citep[see][]{west1997bayesian}. The parameter $\delta_V \in (0,1]$ operates as a discount factor, so the larger the discount is, smaller will be the random shock in the observational covariance. When $\delta_V = 1$, the variance is static over time, i.e., $\sigma^2_t = \sigma^2$, $\forall t$. The initial information at time $t = 0$ assumes $\bftheta_0|\textbf{D}_0 \sim T_{n_0}(\textbf{m}_0,\textbf{C}_0)$ and $\varphi_0|\textbf{D}_0 \sim G(n_0/2,d_0/2)$.

\subsubsection{Spatiotemporal ensemble model output statistics}

Analogously to the model in (\ref{eq:DGOP}), this proposed model combines the Spatial EMOS with DLMs. Also assuming the structure stated in (\ref{eq:BC}), the spatiotemporal Gaussian model for $X_t(s)$ is given by:
\label{eq:STEMOS}
\begin{align}
\textbf{X}_{t}= {\textbf{F}'_{t}}\bftheta_t + \bfvarepsilon_t, \quad & \bfvarepsilon_t \sim N(\textbf{0}_n,\Sigma^*_t), \label{eq:STEMOSxt}\\
\bftheta_t= \textbf{G}_t\bftheta_{t-1} + \bfomega_t, \quad & \bfomega_t \sim N(\textbf{0}_p,\textbf{W}_t), \label{eq:STEMOSthetat}
\end{align}
where $\bfvarepsilon_t = \left( \varepsilon_t(\bfs_1),\dots,\varepsilon_t(\bfs_n)\right)'$ follows a zero mean multivariate normal distribution with covariance matrix $\Sigma_t^*$ with elements $\Sigma_{t_{i,j}}^* =  D_{t_{i,i}}H_{i,j}D_{t_{j,j}},\quad i,j = 1,\dots,n$, also with $H_{i,j} = \exp(-\phi\|{\bfs_i - \bfs_j}\|)$ and $$D_t = \text{diag}\left(\sqrt{\beta_0 + \beta_1 S^2_{1,t}},...,\sqrt{\beta_0 + \beta_1 S^2_{n,t}}\right),$$ an $n$-dimensional diagonal matrix, such that $S^2_{i,t}$ is the sample variance of the ensemble forecast for location $i$ at time $t$. Different from the model in (\ref{eq:DGOP}), $\bfomega_t$ is normally distributed with unknown covariance matrix $\textbf{W}_t$ which can also be estimated using discounting factors. The initial information at time $t = 0$ assumes $\bftheta_0|\textbf{D}_0 \sim N(\textbf{m}_0,\textbf{C}_0)$.

\subsection{Inference procedure}

Let $\textbf{y}=(\textbf{y}_{1},...,\textbf{y}_{T})'$ be the collection of $T$ observed time series at $n$ spatial locations over $S \subset \mathbb{R}^2$ and, $\Theta=(\bftheta_{0:T},\sigma^2_{0:T}, \phi, \lambda)'$
and $\bfTheta^*=(\bftheta_{0:T},\bfbeta,\phi,\lambda)'$ be the parameter vector in (\ref{eq:DGOP}) and (\ref{eq:STEMOS}), respectively,
 such that $\bftheta_{0:T}=(\bftheta_{0},...,\bftheta_{T})$ and $\sigma^2_{0:T}=(\sigma^2_{0},...,\sigma^2_{T})$.

The inference procedure is performed under the Bayesian paradigm and the model specification is complete after assigning a prior distribution to the parameter vectors $\Theta$ and $\Theta^*$. An advantage of following the Bayesian paradigm is that the inference procedure is performed under a single framework with uncertainty about parameter estimation being naturally accounted for. Moreover, uncertainty about spatial interpolations and temporal predictions is naturally described through the credible intervals (CI) of the respective posterior predictive distributions. Assuming that the components of $\Theta$ and $\Theta^*$ are independent {\it a priori}, from Bayes' theorem we obtain the following posterior distribution of $\Theta$:
\begin{equation}
\begin{split}
\label{eq:DGOPposterior}
p(\Theta | \textbf{y}) & \propto p(\textbf{y}|\Theta) \times p(\Theta) \\ 
& \propto \prod_{t=1}^T |\Sigma_t|^{-\frac{1}{2}} \exp\left\lbrace -\frac{1}{2}
\sum_{t=1}^T (\textbf{x}_{t} - \textbf{F}_{t}'\bftheta_{t})'{\Sigma_t}^{-1}(\textbf{x}_{t} - \textbf{F}_{t}'\bftheta_{t})\right\rbrace \\
&\times \exp\left\lbrace -\frac{1}{2}
\sum_{t=1}^T (\bftheta_{t} - \textbf{G}_{t}\bftheta_{t-1})'{\textbf{W}_t}^{-1}(\bftheta_{t} - \textbf{G}_{t}\bftheta_{t-1})\right\rbrace \\
&\times \prod_{\left\lbrace (i,t): y_{t}(s_i)>c \right\rbrace} {\left\lbrace y_{t}(s_i)\right\rbrace}^{\lambda-1} \times p(\Theta).
\end{split}
\end{equation}

The posterior distribution of $\Theta^*$ is obtained analogously. The prior distributions assigned to, respectively, $\Theta$ and $\Theta^*$, are described in Section \ref{subsec:modelset}.

The kernel of this distribution does not result in a known closed-form expression. Markov chain Monte Carlo (MCMC) methods are used to obtain samples from the posterior distribution of interest. In particular, the Gibbs sampling algorithm is used when there are missing or censored observations, i.e., respectively, $Y_t(s)$ is missing or $Y_t(s) \leq c$; the forward filtering backward sampling algorithm \citep[FFBS,][]{fruhwirth1994data,carter1994gibbs} for $\bftheta_{0:T}$ and $\sigma^2_{0:T}$; and the robust adaptive Metropolis algorithm \citep[RAM,][]{vihola2012robust} for the remaining parameters.

% --- Section 4
\section{Application to wind speed forecasts in the state of Minas Gerais}
\label{sec:application}

Following the motivation presented in Sections \ref{sec:intro} and \ref{sec:data}, the proposed postprocessing extensions are used in this section to calibrate hourly wind speed forecasts (24 hours ahead) for the state of Minas Gerais generated by the Eta regional climate model. In particular, we consider a subset of 59 weather stations and a random selection of 20 days per season from the temporal range of the available dataset. Thus, we obtain the overall and per season results in aggregate predictions for 20 days (480 hours) and 80 days (1920 hours), respectively. The results are compared with those obtained through current postprocessing approaches. In this application, we assess the overall performance of the postprocessing approaches in terms of minimizing systematic errors of wind speed forecasts in the study region.

\subsection{Model settings}
\label{subsec:modelset}

\begin{figure}
\begin{center}
\begin{tabular}{cccc}
\makebox{\includegraphics[scale=.4,angle=-90]{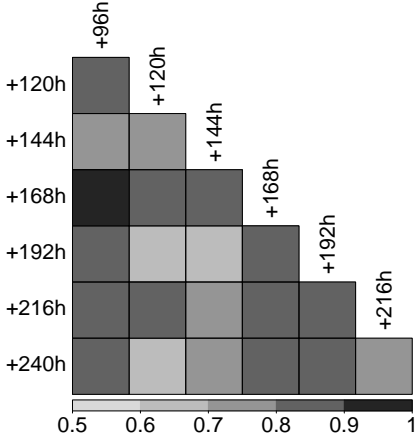}}
\end{tabular}
\end{center}
\caption{Correlation matrix of the ensemble forecast members for wind speed at 10-meter height on 21 June 2016, 1200 UTC. The labels represent the forecasting horizon for each member.}
\label{fig:cormatrix}
\end{figure}

For each fitted model, the same mean structure is defined depending on the mean of the ensemble of wind speed forecasts and a set of auxiliary variables defined at each location: roughness length (see calculation details in Appendix \ref{append:roughness}), latitude, longitude, and elevation. Thus, matrix $\textbf{F}'_{t}$ in (\ref{eq:DGOPxt}) has row components $$ \left(1,\bar{f}(s_i),\text{roughness}(s_i),\text{elevation}(s_i),\text{latitude}(s_i),\text{longitude}(s_i), 1,0 \right)'.$$ The ensemble mean $\bar{f}$ is used to avoid possible multicollinearity problems. This procedure does not entail a great loss of information, since there is a high linear correlation between the members, as shown in Figure \ref{fig:cormatrix}, implying reasonable stability of the numerical forecasts from the Eta model. According to \cite{grimit2007measuring}, the ensemble mean can capture a possible point anomaly (e.g., cold front), so its use is also indicated by theoretical aspects. The seasonal pattern of solar forcing is represented by a Fourier harmonic through evolution matrix $\textbf{G}_t$, which is nested in dynamic linear models. In particular, this matrix is given by $\textbf{G}_t = \text{diag}({\bf G}_{1}, {\bf G}_{2})$, where ${\bf G}_{1}$ is an identity matrix of order 6 and 
$${\bf G}_{2} = \left(
\begin{array}{cc}
{\color{white}+}\cos(2 \pi/24) & \sin(2 \pi/24) \\
-\sin(2 \pi/24) & \cos(2 \pi/24)
\end{array}
\right).$$

To simplify the inference procedure established in Section \ref{subsec:proposed}, the covariance matrix of the evolution equation ${\bf W}_t$ in (\ref{eq:DGOPthetat}) and (\ref{eq:STEMOSthetat}) is estimated using discounting factors \citep{west1997bayesian}. We set the discount factors $\delta_T$ for trend referring to the intercept, the numerical prediction and the geographic location components, and $\delta_S$ for seasonality component. The discount factor $\delta_V$ is exclusive to the DGOP model and works as an artificial instrument to insert random shocks in the observational covariance along time, as shown in (\ref{eq:DGOPpsit}). This analysis considers the exponential correlation function to model spatial dependence. This structure was chosen due to its simplicity. The number of spatial locations in the dataset is not very large ($n=68$). Several papers have indicated that there would not be enough information coming from data to estimate smoothness parameters in this context, such as \cite{zhang2004inconsistent}, which indicates that the parameters in the Mat\'ern model cannot all be estimated consistently. Note, for instance, that if the shortest distance in space is too large, smoothness parameters will be poorly estimated even for large datasets. Given these difficulties for this analysis, we fixed the smoothness and considered the exponential correlation.

Moreover, we assigned reasonably vague priors to model parameters. Specifically, for exclusive parameters of DGOP: $\bftheta_0 \sim T_1(\textbf{0},\textbf{I}_8)$ where ${\bf I}_{8}$ is an identity matrix of order 8 and $\varphi \sim G\left(1, \frac{1}{10}\right)$; for exclusive parameters of STEMOS: $\bftheta_0 \sim N(\textbf{0}_8,\textbf{I}_8)$ and $\bfbeta \sim NT_{(0,\infty)}(\textbf{0}_2,10\textbf{I}_2)$; and for common parameters: $\phi \sim G\left( 2, \frac{\text{max}(d)}{6}\right)$ based on the assumption that the practical range (when the spatial correlation is equal to 0.05) is reached at half of the maximum distance ($\text{max}(d)$/2) between weather station locations and $\lambda \sim N(1,10)$, such that the expected value is 1, representing the null effect of the Box-Cox transformation. The robustness of these prior distributions in simulated experiments is displayed in Appendix \ref{append:robustprior} (Figure \ref{fig:prior}).

\begin{table}
\caption{Summary of features and discount factor setup for fitted models in the calibration process of wind speed forecast at 10-meter height from the Eta model.
}
\label{tab:models}
\centering
\fbox{
\begin{tabular}{lccrrr}
\multicolumn{1}{c}{\multirow{2}{*}{Model}} & 
\multicolumn{2}{c}{Features}               & 
\multicolumn{3}{c}{Discount factors} \\ \cline{2-6} \multicolumn{1}{c}{} & \begin{tabular}[c]{@{}c@{}}Dynamic \\ parameters\end{tabular} & 
\begin{tabular}[c]{@{}c@{}}Spatial \\ component\end{tabular} & 
$\delta_T$ & $\delta_S$ & $\delta_V$ \\ 
\hline
MOS    & No  & No  & 1   & 1   & -   \\
GOP    & No  & Yes & 1   & 1   & 1   \\
SEMOS  & No  & Yes & 1   & 1   & -   \\
DMOS   & Yes & No  & $\tilde{\delta}_T$ & $\tilde{\delta}_S$ & - \\
DGOP   & Yes & Yes & $\tilde{\delta}_T$ & $\tilde{\delta}_S$ & $\tilde{\delta}_V$ \\
STEMOS & Yes & Yes & $\tilde{\delta}_T$ & $\tilde{\delta}_S$ & -      
\end{tabular}}
\end{table}

To evaluate the improvements in the calibration of wind speed forecasts for the state of Minas Gerais, six models were fitted to the dataset described in Section \ref{sec:data}. The models were obtained from distinct settings of the proposed models, which are described in Table \ref{tab:models}. A sensitivity analysis, described in Appendix \ref{append:sensdiscount} (Table \ref{tab:sensresults}), was performed to choose the best combination of discount factors ($\tilde{\delta}_T$, $\tilde{\delta}_S$, $\tilde{\delta}_V)'$ for ($\delta_{T}$, $\delta_{S}$, $\delta_{V})'$. This was used as an alternative to directly estimate this parameter which avoids over-smoothing the observed data and aims to achieve good predictive power.

All setups considered left censoring with Box-Cox transformation as described in (\ref{eq:BC}). Establishing certain parameters of the proposed models on specific values, we obtain simpler models as particular cases. For instance, if the discount factors are set to 1, the corresponding dynamic parameters become static over time. Setting the transformation parameter $\lambda$ equal to 1, the Box-Cox transformation keeps the response in its original scale. Finally, when a large value is assigned to the decay parameter of the exponential correlation function $\phi$, the correlation matrix $H$ in (\ref{eq:DGOPxt}) and (\ref{eq:STEMOSxt}) tends to an identity matrix resulting in spatial independence between observed locations. We used a moving window training dataset of 10 days (240 hours) as a training period. 

\subsection{Computational details}
\label{subsec:computat}

The MCMC algorithm was implemented in the \texttt{R} programming language, version 3.4.1 \citep{teamR}. To speed up this simulation process, some particular functions were implemented in \texttt{C++} language using the library \texttt{Armadillo} \citep{sanderson2016armadillo} through the package \texttt{Rcpp} \citep{eddelbuettel2011rcpp}. The analyses were carried out using a laptop computer with a 2.70 GHz Intel Core i7-7500 processor, 32 GB RAM, running with a Microsoft Windows 7 Professional operating system.

For each fitted model over time, convergence tests were performed to check agreement and convergence of two parallel chains starting from different values. The convergence diagnostic was given through the dependence factor \citep{raftery1995number} and the $\hat{R}$ statistics \citep{gelman1992inference}. With chains convergence ensured by these tests, we ran operationally a single chain with 12,500 iterations, discarding the first 500 and sampling at every 5$^\text{th}$ step. 

% $\textsuperscript{\textregistered}$
% $\textsuperscript{\texttrademark}$

\subsection{Results}
\label{subsec:results}

Model comparison is performed through the out-of-sample predictive performance using the root-mean-square error (RMSE), the mean absolute error (MAE), the index of agreement \citep[$d$,][]{willmott1981validation}, and the interval score \citep[IS,][]{gneiting2007strictly}. The goodness of fit also is determined through the well-known Bayesian criteria: the deviance information criterion \citep[DIC,][]{spiegelhalter2002bayesian} and the logarithm of the pseudo marginal likelihood \citep[LPML,][]{gelfand1996model}. Each of these criteria is described in more detail in Appendix \ref{append:criteria}. The computational efficiency is also used as a secondary model comparison criterion. 

\begin{table}
\caption{Average MAE, RMSE, $d$, IS, DIC, LPML, and computational time (in minutes) of 24-hour deterministic forecasts for wind speed at 10-meter height.}
\label{tab:results}
\centering
\fbox{ 
\begin{tabular}{lrrrrrrr}
\multicolumn{1}{c}{Forecast} & \multicolumn{1}{c}{MAE} & \multicolumn{1}{c}{RMSE} & \multicolumn{1}{c}{$d$} & \multicolumn{1}{c}{IS} & \multicolumn{1}{c}{DIC} & \multicolumn{1}{c}{LPML} & \multicolumn{1}{c}{Time (min)} \\
\hline
Ensemble mean & 1.66 & 1.95 & 0.48 & - & - & - & - \\
\multicolumn{1}{c}{\textit{Static models}} & & & & & & &\\
MOS           & 1.07 & 1.32 & 0.44 & 21.90 & 22939 & -11454 & 37.60 \\
GOP           & 1.07 & 1.33 & 0.42 & 78.74 & 53120 & -18699 & 47.32 \\
SEMOS         & 1.08 & 1.33 & 0.42 & 46.23 & 52312 & -18527 & 66.38 \\
\multicolumn{1}{c}{\textit{Dynamic models}} & & & & & & &\\
DMOS          & 0.94 & 1.17 & 0.57 & 4.10 & 21085 & -10413 & 37.85 \\
DGOP          & 0.94 & 1.19 & 0.58 & 4.38 & 32708 & -12777 & 46.87 \\
STEMOS        & 0.94 & 1.19 & 0.58 & 4.49 & 32519 & -12736 & 66.39
\end{tabular}}
\end{table}

Table \ref{tab:results} reports the average of the MAE, RMSE, $d$, IS, DIC, LPML, and computational time (in minutes) for 24-hour wind speed forecasts at 10-meter height obtained under the Eta model and the six fitted models. All the fitted models present better results than the ensemble mean for all the criteria considered in common. In particular, models considering temporal dynamics performed better than the others according to all predictive performance criteria. Considering the MAE criterion, they were better than ensemble mean and the best static model (MOS), respectively, by 43\% and 13\%. Similar results were obtained in terms of the RMSE. Concerning the index of agreement $d$, while the ensemble mean performed better than the static models, the dynamic models outperformed the ensemble mean by 20\% and the static models by 32\%. Thus, considering the prediction intervals of probabilistic forecasts, the static models obtained the highest values of IS. Moreover, the IS criterion allowed identifying clearer differences between the dynamic models than the other criteria considered. The models that considered the purely spatial component performed significantly worse than the rest in terms of the DIC and LPML. In contrast, the addition of the temporal component benefited both predictive performance and goodness of fit. Remarkably, the DMOS presented better results for all criteria.

Moreover, the boxplots of the average MAE, RMSE and $d$ of 24-hour deterministic forecasts for wind speed at 10-meter height during the seasons are provided as additional figures in Appendix \ref{append:sup} (Figures \ref{fig:MAE} -- \ref{fig:d}). It shows that the forecast evaluation criteria considered, along with wind speed, have different patterns across seasons. Regarding computational time, the DGOP-based models were more efficient on average, taking about 2.4\% to 8.9\% less time than STEMOS-based models. Its better performance is related to the sequential estimation of dynamic parameters.

\begin{figure}
 \centering
 \makebox{\includegraphics[scale=.5, angle=-90]{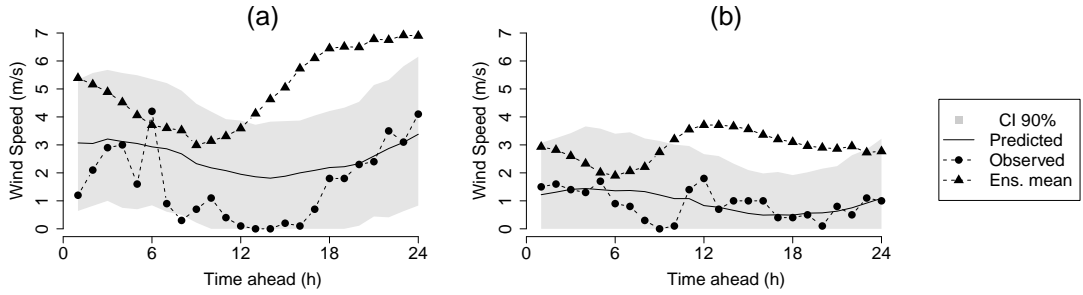}}
 \caption{DGOP 24-hour probabilistic forecast window for wind speed at 10-meter height for Vi\c cosa weather station from (a) 08 November 2016, 1300 UTC to 09 November 2016, 1200 UTC and (b) 12 April 2017, 1300 UTC to 13 April 2017, 1200 UTC. The solid circles represent the numerical prediction. The solid triangles represent the actual values, the solid line represents the bias-corrected deterministic forecasts, and the shaded area represents the 90\% prediction interval.}
\label{fig:plocalA510}
\end{figure}

\begin{figure}
\centering
\makebox{\includegraphics[scale=.65, angle=0]{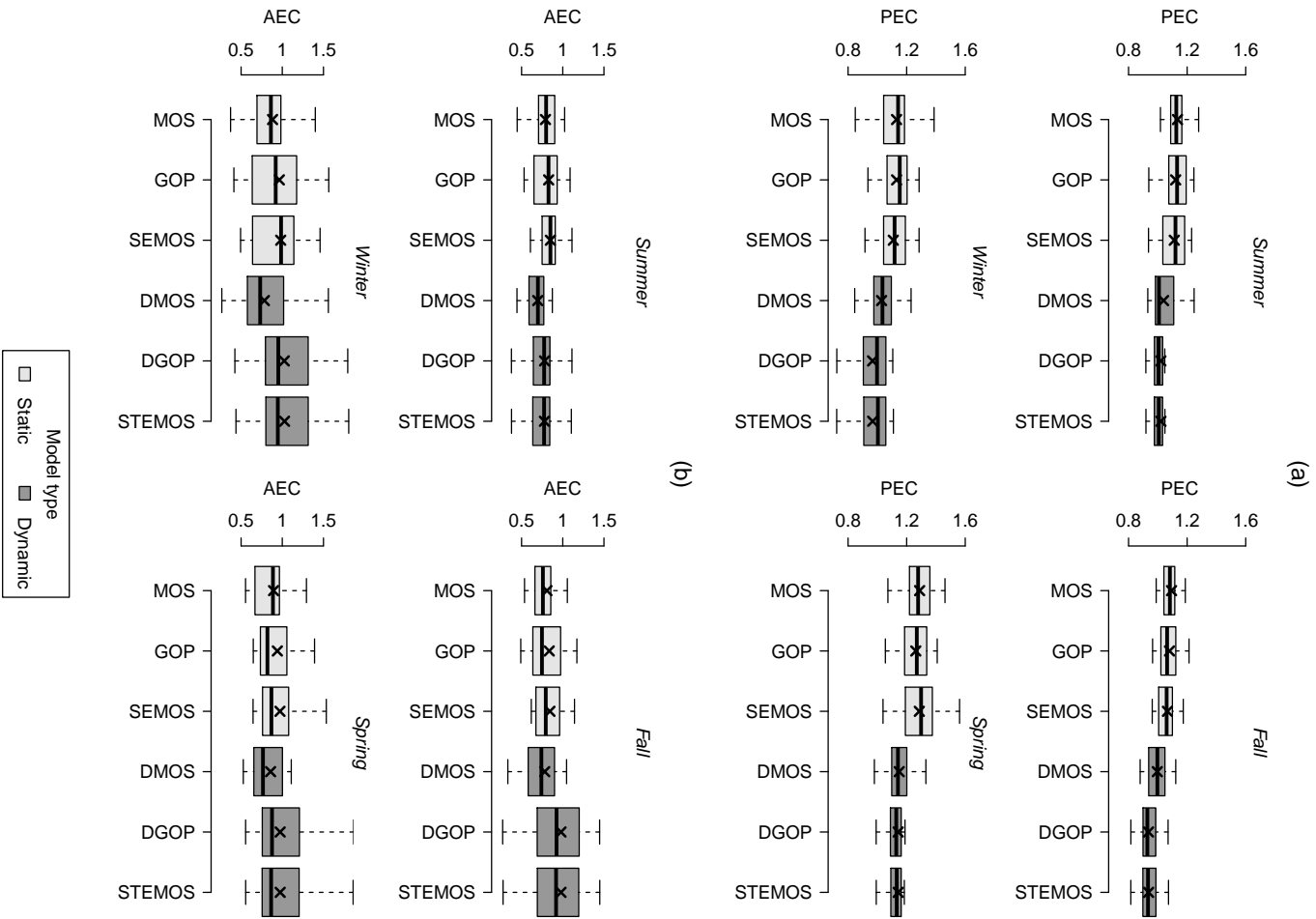}}
  \caption{Boxplots of average decomposition of RMSE: (a) phase error component (PEC) and (b) amplitude error component (AEC) of 24-hour deterministic forecasts for wind speed at 10-meter height during the seasons. The dense horizontal line and the $\mathbf{\times}$ symbol in each boxplot represent respectively, the median and the mean values.}
\label{fig:RMSEdecomp}
\end{figure}

Among the model comparison criteria considered, the RMSE has a particular feature that can be used to discriminate different error sources in the context of calibrating wind speed forecasts. In general, error sources are related to either local conditions or general properties of the NWP models. According to \cite{lange2005uncertainty}, this criterion can be decomposed into two additive parts: the amplitude and the phase errors. Figure \ref{fig:plocalA510} shows illustrative panels for each of these additive components recorded at Vi\c cosa weather station. In some cases, these components may be not explicitly well-defined generating clear visualization. Generally, both occur but with different intensities. The amplitude errors are related to forecasts with correct temporal evolution but with a systematic difference from the actual measure. An observation of this error type is shown in Figure \ref{fig:plocalA510}(a). 
Note the similarity of the trajectory between both predictions and actual values, except for the level. On the other hand, the phase errors are related to forecasts with correct amplitude but with a mismatch of the temporal evolution of the actual measure. Figure \ref{fig:plocalA510}(b) shows a case for which the numerical forecast does not respect the duration of the intraday seasonal cycle. This can be considered a phase error. Specific corrections are indicated for each component of the error. Statistical postprocessing methods apply linear corrections and can minimize amplitude errors. However, this process is ineffective for phase errors, which are associated with cross-correlation between the forecast and actual time series. Thus, this error type is invariant under linear transformations. Panels of Figure \ref{fig:RMSEdecomp} display the decomposition components of RMSE for deterministic forecasts from probabilistic models during the seasons. Figure \ref{fig:RMSEdecomp}(a) shows the portion of RMSE associated with phase errors. For all seasons, the forecast errors of dynamic models present lower values for this component than the static models. Among the dynamic models, the full-featured models associate their RMSE values slightly less to phase errors than the simplest model DMOS. In this way, since the decomposition elements of RMSE are complementary, the models DGOP and STEMOS associate greater parts of the RMSE value to the amplitude errors compared to the other models, as shown in Figure \ref{fig:RMSEdecomp}(b).

A larger portion of RMSE associated with amplitude errors still allows substantial improvements through linear corrections. Therefore, the dynamic models have greater potential for better calibrated probabilistic predictions among all fitted models. These benefits can be obtained by including more covariate information about the local pattern of the weather phenomenon or more detailed geographical features of relief (e.g., climatology, vegetation type, and presence of water bodies).

Furthermore, our dynamic proposal described in (\ref{eq:DGOPpsit}) assumes the dynamic variance in time. Therefore, this feature induces fat tails of residuals distribution. In that case, the integrated process is no longer Gaussian but Student-t distributed. Also, Q-Q plots of residuals are presented in Appendix \ref{append:sup} (Figure \ref{fig:qqplot}) as an indication of the good fit of the assumed sampling distribution. The STEMOS presents better results in terms of tail modeling, indicating that this model is able to better capture large observations compared to the other fitted models.

\begin{figure}
 \centering
 \makebox{\includegraphics[scale=.5, angle=-90]{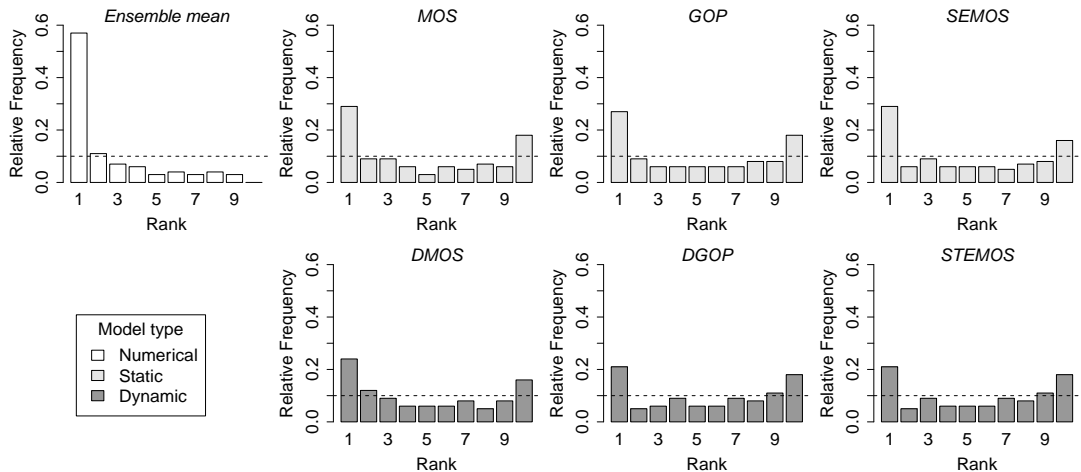}}
 \caption{Verification rank histograms of forecasts for wind speed at 10-meter height in Minas Gerais.}
 \label{fig:rankhist}
\end{figure}

An overall calibration assessment of forecasts was carried out through the verification rank histogram \citep{anderson1996method} which evaluates the forecast reliability. An ideal calibration is obtained when its bins are uniformly distributed. Figure \ref{fig:rankhist} shows the verification rank histograms for forecasts from all fitted models considering the aggregate result for all periods and weather stations in Minas Gerais. The skewed aspect of the histogram for forecasts obtained by the ensemble mean implies that the measurements at weather stations are lower than the predictions, i.e., the ensemble mean commonly overestimates the actual values. All tested calibration models significantly improve the numerical forecasts. Specifically, the forecasts from statistical postprocessing models present similar histograms. These histograms have aspects closer to a uniform distribution than the histogram for uncalibrated forecasts. This result indicates greater reliability. Indeed, the calibration process removes the systematic error resulting in more accurate forecasts as shown in Table \ref{tab:results}. The forecasts from dynamic models have slightly less underdispersed distribution. Among these, the DGOP and the STEMOS present the best calibration overall. Nevertheless, the observed $\cup$-shape can represent a lack of variability. The irregularities of boundary bins (bins 1 and 10) suggest that the calibrated forecasts are not ideal at the outermost values. According to wind speed measurements at all weather stations in Minas Gerais, we interpret the outermost values 0.2 and 3.8, respectively, as 0.1 and 0.9-quantiles.
% (0.0 and 4.6, respectively 0.05 and 0.95-quantiles).

\begin{figure}
\centering
\makebox{\includegraphics[scale=.6, angle=-90]{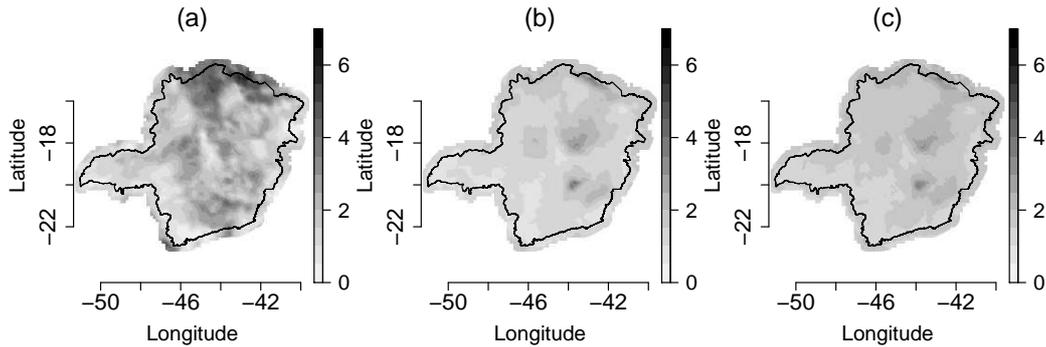}}
\caption{24-hour weather field forecast of wind speed at 10-meter height in the state of Minas Gerais on 21 July 2016, 1200 UTC: (a) Numerical prediction, (b) DGOP deterministic forecast, and (c) DGOP margin of error defined as half the width of the 90\% prediction interval.}
\label{fig:spatpred}
\end{figure}

The models with spatial component (see Table \ref{tab:models}) allow straightforward spatial interpolation through kriging \citep{cressie1993statistics}, assuming that the latent process $X_t(\bfs)$ in (\ref{eq:BC}) is a realization of the Gaussian random fields. Therefore, potential improvements in site forecasts may be obtained from the proposed statistical postprocessing processes over the entire weather field. Figure \ref{fig:spatpred} shows the 24-hour weather field forecast of wind speed at 10-meter height in Minas Gerais starting on 21 July 2016, 1200 UTC. Figure \ref{fig:spatpred}(a) shows the numerical prediction for the entire weather field obtained using the bilinearly interpolated mean of ensemble forecasts. Figure \ref{fig:spatpred}(b) shows the deterministic DGOP weather field forecast defined as the median of the predictive distribution. Note that the calibrated forecasts are smoothly spatially distributed in contrast to the numerical prediction. As shown in Figure \ref{fig:rankhist}, the ensemble members from the Eta model, specifically the ensemble mean, generally overestimate the actual wind speed at 10-meter height. Thus, bias-corrected smooth distribution over the weather field has greater reliability. Lastly, Figure \ref{fig:spatpred}(c) shows the margin of error of the 90\% prediction interval defined as half the width of the credible interval. This plot is useful to indicate regions where there is more uncertainty of predictions. Three regions on the map, closer to the center, are spotted as having larger variability.

\begin{figure}
\centering
\makebox{\includegraphics[scale=.55, angle=-90]{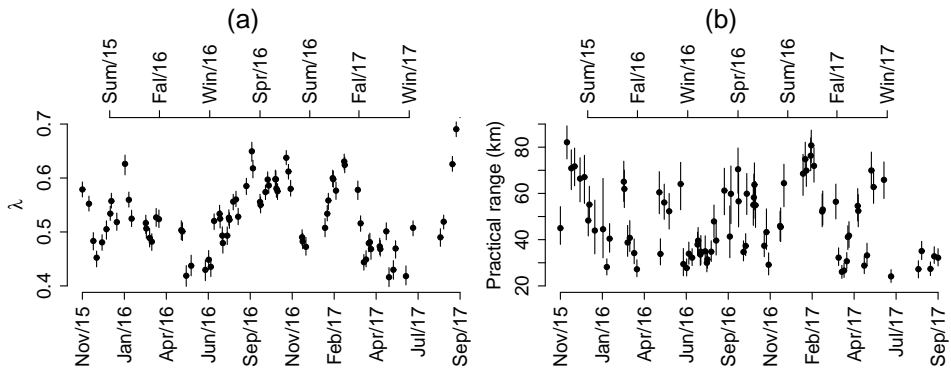}}
\caption{Posterior distribution over the daily range of (a) the Box-Cox transformation parameter $\lambda$, and of (b) the decay parameter $\phi$. The solid circle represents the posterior median and the vertical line represents the 90\% CI.}
\label{fig:postparam}
\end{figure}

Regarding model parameter fitting and interpretation, Figure \ref{fig:postparam} shows the posterior distribution behavior of selected parameters from DGOP. Figure \ref{fig:postparam}(a) shows the posterior distribution of the Box-Cox transformation parameter $\lambda$ over the daily range. The estimated value of this parameter ranges between 0.4 and 0.7. The log transformation ($\lambda=0$), commonly indicated for wind speed data normalization \citep{zhu2012short}, is not supported by this range. Although the square root transformation ($\lambda=0.5$), also indicated, is supported, this parameter performs a cyclical movement related to the seasons. For the summer and spring, which have the highest wind speeds, higher values of $\lambda$ are suggested. The opposite occurs for seasons with low wind profile (fall and winter). As expected, the behavior of wind speed data is less skewed in fields where the records have a higher level \citep[see][]{gel2004calibrated}. Therefore, values closer to 1 are suggested for the Box-Cox transformation parameter, implying a weak or null ($\lambda=1$) effect for this transformation. For this application, these conditions are not supported by the estimated range at any time. These results indicate that the data transformation should not be restricted to a single and static transformation when the temporal windows are wide. If the model is going to be fitted for a whole year, then a $\lambda_t$ process might be needed. In our case, it is sufficient to assume a constant $\lambda$ over the training period of 10 days (240 hours). Figure \ref{fig:postparam}(b) shows the posterior distribution of the spatial correlation decay parameter $\phi$ over the daily range. The practical range represents the distance where the spatial correlation between the weather stations decays until almost zero. For example, this distance is estimated to range between 20 and 90 km. In practice, the weather stations within an approximate 2 to 6 grid cell radius are correlated with each other since the Eta model grid has a resolution of 15 km (see Figure \ref{fig:locations}). This parameter also performs a cyclic movement. These values are evidence that the decay parameter can vary over seasons. If a wide temporal window is considered then $\phi_t$ can be estimated. Note that this approach would be computationally demanding since the precision matrices would be inverted at each time point. Since our aim is the practical use of the proposed models in the context of wind power prediction, we did not pursue this avenue further. Moreover, the parameter $\phi$ is well estimated only if the spatial correlation decreases to (approximately) zero in the observed spatial domain. If this is not the case for cross-sectional data, then temporal observations may aid in the adequate estimation of this parameter. Thus, fixing $\phi$ to be the same for each station in our analysis resulted in a consistent estimation of this parameter and satisfactory interpolations.
 
% --- Section 5
\section{Application to simulated experiments}
\label{sec:applicationsimul}

Following the results for hourly calibration of wind speed forecasts in the state of Minas Gerais presented in Section \ref{sec:application}, in this section we apply the proposed postprocessing extensions to evaluate the improvements that can be produced under different conditions in simulated experiments. The simulated weather fields have similar characteristics to the real data application: only non-negative values are allowed with left censoring, and skew behavior. The extension of the simulated region is also considered, preserving approximately the same distance between the measurement locations.

We designate the simulated experiments as scenarios. The three scenarios are implemented as follows. The first scenario is a simplistic representation of a weather field generated by the MOS model. This experiment aims to explore the use of full-featured models when an elementary field is available. Although these methods are not parsimonious in this context, we explore how the complexity of the models can impact the calibration process and its results in practice. The remaining scenarios are compound, involving temporal and spatial statistical components, and are generated by DGOP and STEMOS models, respectively. The difference between these experiments is the same as that between the generating models, i.e., the latter accounts for the spread-skill relationship. Furthermore, we investigate the loss of calibration performance when a simpler method is applied in situations where more robust methods are required. In this way, we intend to validate and encourage the use of the proposed postprocessing extensions in applications ranging from simple to challenging. The experiments use the same model settings already considered including the mean structure of the models, the prior distributions, and the training period of 10 days (240 hours). The calibration performance is evaluated as before (see Table \ref{tab:results}). 

The average of the MAE, RMSE, $d$, IS, DIC, and LPML for 24-hour simulated wind speed forecasts are shown in Table \ref{tab:simulated_results}. All experiments were generated from a fixed simulated ensemble. The difference between the results obtained by the ensemble mean is related to different structures of the generating models in each scenario. We expected that these results to be similar but not necessarily identical. Scenario 1 is the simplest one regarding reduction of systematic errors according to the comparison criteria. Note that a comprehensible field facilitates the calibration performed by the models. The MOS model, generator of this scenario, obtained the best criteria regarding the adjustment to the data (DIC and LPML), as expected. However, its direct extensions, GOP, DMOS, and DGOP, resulted in similar values. These results confirm the good ability of the proposed postprocessing extensions to adapt parameters to limit values when necessary. The ability to nest precursor models proved to be functional. Thus, the operational use of our extensions is encouraged independently of the weather field characteristics. The gains of scenarios 2 and 3 are smaller compared to the previous scenario due to the greater complexity involved in this simulation. Given these conditions, the performance improved with greater number of featuresin the model. In both scenarios, each proposed extension, DGOP and STEMOS, obtained the best results even when it was not the generating model. For example, these gains ranged from 48\% in scenario 3 to 54\%, in scenario 2 concerning the ensemble mean for the MAE criterion. The implementation of the MOS model in scenario 3 resulted in worse performance than the ensemble mean in this same criterion, defeating the purpose of postprocessing models. From these experiments, the importance of choosing the model to apply in a given weather field is evident. Applying a simple model in complex scenarios can result in a poor gain or even loss in the calibration process, as seen in scenario 3, while the opposite does not occur, as seen in scenario 1.

% Scenario 3b - underdispersed
% Scenario 4 - overdispersed

\begin{table}[ht]
\caption{Average MAE, RMSE, $d$, IS, DIC and LPML of 24-hour deterministic forecasts in simulated experiments.}
\label{tab:simulated_results}
\centering
\fbox{ 
  \begin{tabular}{lrrrrrr}
  \multicolumn{1}{c}{Forecast} & \multicolumn{1}{c}{MAE} & \multicolumn{1}{c}{RMSE} & 
    \multicolumn{1}{c}{$d$} & \multicolumn{1}{c}{IS} & \multicolumn{1}{c}{DIC} & 
    \multicolumn{1}{c}{LPML} \\ \hline
  \multicolumn{1}{c}{\textit{Scenario 1}} &       &       &       &       &       &        \\
  Ensemble mean       & 3.405 & 3.544 & 0.382 & -     & -     & -      \\
  MOS                 & 0.752 & 0.941 & 0.709 & 3.126 & 12961 & -6472  \\
  GOP                 & 0.753 & 0.942 & 0.710 & 3.156 & 13212 & -6691  \\
  SEMOS               & 0.753 & 0.942 & 0.710 & 3.172 & 13740 & -6797  \\
  DMOS                & 0.752 & 0.941 & 0.708 & 3.146 & 13059 & -6490  \\
  DGOP                & 0.754 & 0.943 & 0.709 & 3.182 & 13355 & -6715  \\
  STEMOS              & 0.753 & 0.942 & 0.709 & 3.178 & 13715 & -6787  \\
  \multicolumn{1}{c}{\textit{Scenario 2}} &       &       &       &       &       &        \\
  Ensemble mean       & 3.405 & 3.781 & 0.471 & -     & -     & -      \\
  MOS                 & 2.942 & 3.407 & 0.504 & 8.917 & 41934 & -20975 \\
  GOP                 & 2.839 & 3.312 & 0.510 & 8.862 & 33863 & -16619 \\
  SEMOS               & 2.777 & 3.251 & 0.514 & 8.726 & 35288 & -16972 \\
  DMOS                & 1.626 & 1.988 & 0.590 & 9.493 & 28360 & -14141 \\
  DGOP                & 1.590 & 1.947 & 0.602 & 9.135 & 27439 & -13727 \\
  STEMOS              & 1.554 & 1.910 & 0.596 & 5.413 & 29155 & -14605 \\
  \multicolumn{1}{c}{\textit{Scenario 3}} &       &       &       &       &       &        \\
  Ensemble mean       & 3.059 & 3.429 & 0.504 & -     & -     & -      \\
  MOS                 & 3.178 & 3.645 & 0.507 & 9.841 & 45318 & -22676 \\
  GOP                 & 3.020 & 3.497 & 0.518 & 9.710 & 36131 & -17759 \\
  SEMOS               & 3.013 & 3.487 & 0.519 & 8.877 & 36700 & -17934 \\
  DMOS                & 1.617 & 1.996 & 0.658 & 6.777 & 33170 & -16570 \\
  DGOP                & 1.562 & 1.922 & 0.646 & 6.240 & 31572 & -15775 \\
  STEMOS              & 1.587 & 1.934 & 0.647 & 9.371 & 29542 & -14808
  \end{tabular}}
\end{table}

% --- Section 6
\section{Discussion}
\label{sec:discussion}

This work presents two new statistical calibration models for weather fields forecasts obtained from regional climate models. The proposed models generalize the well-established spatial postprocessing techniques GOP and SEMOS by combining them with Bayesian dynamic models. Thus, the proposed methods take into account the spatial and temporal correlations, allowing the spatiotemporal calibration of forecasts obtained from NWP models. 

As discussed in Section \ref{sec:intro}, the homogenization of topographic relief achieved by the NWP models results in outputs with potential spatially correlated errors. Thus, the calibration through geostatistical models preserves the inherent spatial correlation structure of the weather field. Additionally, the intrinsic seasonal patterns of meteorological phenomena, discussed in Section \ref{sec:postmodels}, motivate the use of postprocessing techniques that include temporal correlation in the calibration process as the Bayesian dynamic models do. In contrast with the well-established calibration methods that result in a single forecast for a fixed horizon, the proposed methods allow a single sequential calibration over time, more suitable to hourly data, producing a bias-corrected forecast window output. Moreover, the dynamic structure avoids the issue of empirically optimizing the length of training period due to the intrinsic ability to weight past observations over time \citep[see][]{west1997bayesian}. As \cite{raftery2005using} suggested, the proposed methods provide an automatic way of choosing the length of the training period. This is balanced such that, parameter estimation is not affected starting from a minimum length threshold, thus avoiding distortions of the calibration process. Also, other advantages of our methods are the natural manipulation of missing data and the versatility of models.
Through the data augmentation technique, the missing data are interpreted as latent data and parameter estimation is performed by a fully Bayesian approach, without a large computational cost.

Our proposed model assumes the Box-Cox transformation parameter $\lambda$ to be fixed over time and, for our application, this is reasonable for each season. However, if the model is going to be used for the whole year, a moving $\lambda_t$ might provide better predictions, since asymmetry can vary with the seasons. This issue is a subject of future research.

In particular, we investigate the calibration process of wind forecasts at 10-meter height, which are well known to be locally predicted with systematic errors. As discussed in Section \ref{sec:data}, these errors are associated with the complex landscape of Minas Gerais and its strong influence on wind speed behavior at low heights. The study region can be considered a low wind speed area. Figure \ref{fig:data} shows a common pattern recorded at the 68 weather stations: a large number of zero observations, rare wind speeds above 6 m/s, skewed distribution, and different wind speed behavior from other seasons during spring. These characteristics are also commonly observed in many precipitation modeling and forecasting applications. In line with the application performed by \cite{sanso1999venezuelan}, the proposals introduce data transformation within the dynamic model, resulting in a flexible sampling distribution for the errors, which can potentially be asymmetric. Hence, our approach leads to narrower predictions in comparison with simpler models without temporal dependence. In this application, the space-time models overall result in slightly better calibration performance, as shown in Table \ref{tab:results} of Section \ref{subsec:results}. The decomposition of the RMSE, explored also in Figure \ref{fig:RMSEdecomp}, shows that the proposed models associate greater portions of their RMSE with amplitude errors that can still be corrected through linear transformations. This association implies that potential improvements can still be made by including additional information using auxiliary variables. Among the probabilistic models, DGOP and STEMOS produced more reliable forecasts, as shown in Figure \ref{fig:rankhist}. The persistence of $\cup$-shape of verification rank histograms indicates a lack of variability. However, the specific irregularities in boundary bins can also be an indicator of conditional biases. \cite{hamill2001interpretation} suggested a local exploratory analysis of model fit and forecast variability. Given the large variability of Minas Gerais relief, the simplistic geostatistical models (see Table \ref{tab:models}) may also underestimate local characteristics. The exponential correlation function was used due to its good functionality and applicability in the calibration context, as seen in \cite{gel2004calibrated}, \cite{berrocal2007combining} and \cite{feldmann2015spatial}. More general correlation functions were not considered due to the difficulty of estimating the smoothness parameters, as discussed in \cite{zhang2004inconsistent}.

The inference procedure was carried out through a fully Bayesian approach. The required prior distributions assigned for model specification were weakly informative. This approach has attractive features such as naturally taking into account the uncertainty about parameter estimation, but it is still expensive computationally as shown in Table \ref{tab:results}. Therefore, it is only recommended for medium-term and long-term forecasts. 

Finally, useful structural improvements can be achieved, for example, by weighting each available ensemble member, as proposed by \cite{scheuerer2015probabilistic}. We work with a particular case in which the same weight is assigned to each member, i.e., the ensemble (arithmetic) mean. \cite{feldmann2015spatial} also proposed the inclusion of local climatological information through auxiliary variables. The proposed models can also be applied to space-time calibration of forecasts from grid-based NWP of other asymmetric censored meteorological variables in which their regime influences seasonal patterns (e.g., precipitation). Additional research on this subject is currently underway.
 
% --- Acknowledgements
\section*{Acknowledgements}
We thank Dr. Chou Sin Chan (CPTEC/INPE) for providing the Eta model outputs used in this study and for her advice during the research. The first author also thanks the financial support from the partnership between Funda\c{c}\~{a}o de Amparo \`{a} Pesquisa do Estado de Minas Gerais (FAPEMIG) and Companhia Energ\'{e}tica de Minas Gerais S.A. (CEMIG) under Project APQ-03813-12.

% --- Appendices
\appendix 
\addcontentsline{toc}{section}{Appendices}

\section{Calculation of the roughness length} 
\label{append:roughness}
According to \cite{hansen1993surface}, the vegetation present on a surface influences the aerodynamic roughness characteristics encountered by the mean wind flow over that surface, affecting both the mean wind speed and direction predicted by numerical models and various other atmospheric parameters. Therefore, the surface roughness length, $z_0$, defined as the height at which the wind speed equals zero, has an important role in the modeling of atmospheric processes.

The aerodynamic roughness length parameter, $z_0$, used as a covariate in the application, was estimated for each calibration site from key atmospheric variables, following some principles of the Monin-Obukhov similarity theory \citep{monin1954basic}. Particularly, assuming a logarithmic wind profile, the averaged wind speed $\overline{u_i}$ at height $z_i$ (10 meters), given by:

\begin{equation*}
\overline{u_i}= \frac{u_{*}}{k} \left[ \ln \left( \frac{z_i}{z_{0}}\right) -\Psi(\zeta_i) \right]
\label{eq:perfil1}
\end{equation*}
was used to derive the roughness length $z_0$ as 

\begin{equation}
z_{0}= z_{i}\exp \left(\frac {-\overline{u_{i}} k}{{u}_{*}} - \Psi(\zeta_{i})\right),
\label{eq:z02}
\end{equation}
where $k$ is the von Karman constant, $u_{*}$ is the friction velocity, $\Psi(\zeta_{i})$ is the stability correction function of the wind profile and $\zeta_i=\frac{z_i}{L}$ is the dimensionless stability parameter given by the height above ground, $z_i$, normalized by the Obukhov length, $L$.

Measurements of air temperature, air pressure, sensible heat flux, momentum flux, and wind stress can be used to derive the parameters $u_{*}$, $L$ and $\Psi(\zeta_{i})$. In this application hourly air temperature and air pressure data were obtained from the available meteorological stations, while hourly reanalysis data from the CFSV2 model \citep{saha2014ncep} were used as the source of heat and momentum fluxes, after interpolation from the CFSV2 regular grid to the calibration sites.

Hourly values of the roughness length, $z_0$, were first obtained from (\ref{eq:z02}) for the two years of available data. Then, median values of $z_0$ by month and by hour within each month (288 values in total for each calibration site) were calculated, considering only those $z_0$ values that were estimated during neutral conditions of atmospheric stability (such condition is achieved when $|L|>500$).

\newpage

\section{Robustness of prior distributions}
\label{append:robustprior}

\begin{figure}[ht]
 \centering
 \makebox{\includegraphics[scale=.65, angle=-90]{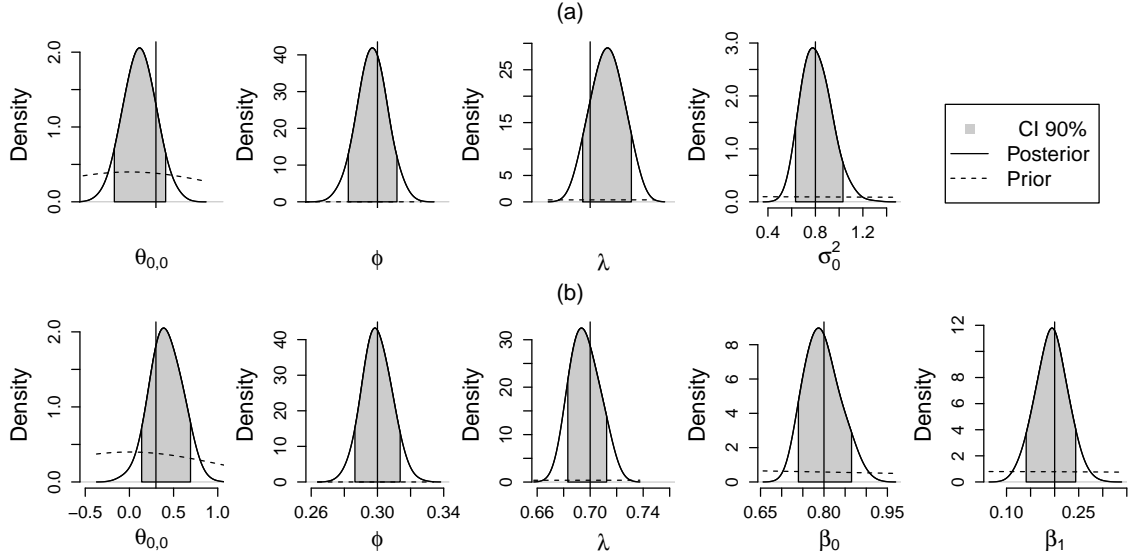}}
  \caption{Posterior and prior distributions associated with parameters of (a) DGOP and (b) STEMOS in simulated experiments. The vertical line represents the actual value assigned for parameters.}
\label{fig:prior}
\end{figure}

In this section, we verify the robustness of prior distributions assigned to the parameters of the proposed postprocessing models applied in Section \ref{sec:application}. Based on these prior distributions in simulated experiments, the posterior and prior distributions associated with the parameter vectors from DGOP and STEMOS are presented in Figure \ref{fig:prior}. For clarity of exposition, only the results of a subset of common parameter $\bftheta_0$ (vector with dimension $r$) are exhibited. Specifically, the intercept parameter at moment $t=0$ represented by $\theta_{0,0}$ is shown. 

Even when using this specific set of prior distributions in a distinct application from that previously reported, Figure \ref{fig:prior} shows that the posterior distributions differ significantly from the prior distributions for all parameters. This is evidence that the gain of information is mostly provided by the data, ensuring that non-informative properties are preserved in general applications.

\newpage

\section{Sensitivity analysis of discount factors}
\label{append:sensdiscount}

This sensitivity analysis was performed to choose the best combination of discount factors ($\tilde{\delta}_T$, $\tilde{\delta}_S$, $\tilde{\delta}_V)'$ for ($\delta_{T}$, $\delta_{S}$, $\delta_{V})'$. In particular, we also consider a subset of 59 weather stations and a random selection of 5 days per season from the temporal range of the available dataset. Following the same model comparison criteria described in Section \ref{sec:application}, Table \ref{tab:sensresults} reports the sensitivity analysis of discount factors through average MAE, RMSE, $d$, IS, DIC, and LPML values for 24-hour wind speed forecasts at 10-meter height. The setups of the dynamic models differ only in the discount factor values. The results indicate greater sensitivity for $\delta_T$, which significantly reduces the predictive performance when assuming lower values. In contrast, the variation of values for $\delta_{S}$ and $\delta_{V}$ only slightly alters the predictive performance. Thus, we determine ($\tilde{\delta}_T$, $\tilde{\delta}_S$, $\tilde{\delta}_V)' = (0.99, 0.95, 0.99)'$.

\begin{table}[ht]
\caption{Average MAE, RMSE, $d$, IS, DIC, and LPML of 24-hour deterministic forecasts for wind speed at 10-meter height (sensitivity analysis of discount factors).}
\label{tab:sensresults}
\centering
\fbox{ 
\begin{tabular}{lrrrrrrrr}
\multicolumn{1}{c}{\multirow{2}{*}{Model}} & \multicolumn{3}{c}{Discount factors}                                                             & \multicolumn{5}{c}{Comparison criteria}                                                                                           \\ \cline{2-9} 
\multicolumn{1}{c}{}                       & \multicolumn{1}{c}{$\delta_T$} & \multicolumn{1}{c}{$\delta_S$} & 
\multicolumn{1}{c}{$\delta_V$} & \multicolumn{1}{c}{MAE} & \multicolumn{1}{c}{RMSE} & \multicolumn{1}{c}{$d$} & 
\multicolumn{1}{c}{DIC} & \multicolumn{1}{c}{LPML} \\ \hline
\multicolumn{1}{c}{\textit{Dynamic models}} & & & & & & & &\\
DMOS & & & & & & & & \\
$\quad$ Setup A & 0.99  & 0.99 & -    & 0.93 & 1.25 & 0.63 & 20967  & -10358 \\
$\quad$ Setup B & 0.99  & 0.95 & -    & 0.94 & 1.25 & 0.63 & 20998  & -10339 \\
$\quad$ Setup C & 0.99  & 0.9  & -    & 0.94 & 1.26 & 0.63 & 21043  & -10318 \\
$\quad$ Setup D & 0.95  & 0.99 & -    & 0.94 & 1.27 & 0.61 & 21226  & -10311 \\
$\quad$ Setup E & 0.9   & 0.99 & -    & 0.95 & 1.28 & 0.61 & 21984  & -10425 \\
DGOP & & & & & & & & \\
$\quad$ Setup A & 0.99  & 0.99 & 0.99 & 0.94 & 1.29 & 0.62 & 34775  & -13270 \\
$\quad$ Setup B & 0.99  & 0.95 & 0.99 & 0.93 & 1.29 & 0.63 & 36310  & -13645 \\
$\quad$ Setup C & 0.99  & 0.99 & 0.9  & 0.94 & 1.30 & 0.62 & 36394  & -13658 \\
$\quad$ Setup D & 0.99  & 0.95 & 0.9  & 0.94 & 1.29 & 0.63 & 37845  & -14012 \\
$\quad$ Setup E & 0.95  & 0.99 & 0.99 & 0.96 & 1.31 & 0.62 & 65624  & -20840 \\
$\quad$ Setup F & 0.9   & 0.99 & 0.99 & 2.07 & 4.40 & 0.26 & 115777 & -33304 \\
STEMOS & & & & & & & & \\
$\quad$ Setup A & 0.99  & 0.99 & -    & 0.94 & 1.29 & 0.62 & 35483  & -13454 \\
$\quad$ Setup B & 0.99  & 0.95 & -    & 0.93 & 1.29 & 0.63 & 37222  & -13877 \\
$\quad$ Setup C & 0.95  & 0.99 & -    & 0.96 & 1.30 & 0.62 & 69226  & -21744 \\
$\quad$ Setup D & 0.9   & 0.99 & -    & 2.41 & 4.73 & 0.23 & 124234 & -35418 
\end{tabular}}
\end{table}

\newpage

\section{Model comparison criteria} 
\label{append:criteria}
In this section, we briefly describe the model comparison criteria used to compare the prediction of the fitted models in Section \ref{sec:application}. The first three criteria (RMSE, MAE and index of agreement) are appropriate to compare numerical predictions from the Eta model, which provides only deterministic estimates, with the proposed postprocessing models. The probabilistic forecasts are evaluated through IS, which takes into account the amplitude and coverage of the prediction intervals in a parsimonious way.

\subsection{Mean absolute error and root-mean-square error}
Standard measures of goodness of fit were also entertained in this
study for comparison purposes. The root-mean-square error
(RMSE) and the mean absolute error (MAE) are given by:
$$
\mbox{RMSE}=\frac{1}{nT}\sum_{i=1}^{n}\sum_{t=1}^{T}(y_{t}(s_i)-\hat{y}_{t}(s_i))^2
\ \ \ \mbox{and} \ \ \
\mbox{MAE}=\frac{1}{nT}\sum_{i=1}^{n}\sum_{t=1}^{T}|y_{t}(s_i)-\hat{y}_{t}(s_i)|,
$$
where $\hat{y}_{t}(s_i)$ is obtained through a Monte Carlo estimate of the posterior mean of the predictive distribution, that is, $E\left[y_{t}({\bf s}_i) \mid {\bf y}\right]$, across $N$ draws. Smaller values of RMSE and MAE indicate better model fit.

\subsection{Index of agreement} 
\cite{willmott1981validation} introduced a standard measure for assessing the quality of forecasts. The index of agreement ($d$) ranges between 0 (absence of agreement) and 1 (perfect agreement), and is given by:
$$d = 1 - \frac{\sum_{i=1}^n \sum_{t=1}^{T}\left(y_{t}(s_i) - \hat{y}_t(s_i)\right)^2}{\sum_{i=1}^n \sum_{t=1}^{T}\left(|\hat{y}_t(s_i) - \bar{y}|+|y_t(s_i) - \bar{y}|\right)^2},$$
where $\bar{y} = \frac{1}{n} \sum_{i=1}^n\sum_{t=1}^{T} y_t(s_i)$.

\subsection{Interval score}
The interval score \citep[IS,][]{gneiting2007strictly} is a scoring rule for interval predictions considering the symmetric prediction interval with level $(1-\alpha)\times$100\%. The score is rewarded by accurate intervals and penalized when there is no coverage of the forecast. If actual values are contained in the prediction interval, this measure is reduced to the range amplitude. The average IS is given by:
\begin{equation*}
\begin{split}
\text{IS} = & \quad \frac{1}{nT}\sum_{i=1}^n\sum_{i=1}^T (\hat{u}_t(s_i)-\hat{l}_t(s_i)) \\
 & +\frac{2}{\alpha}(\hat{l}_t(s_i)-y_t(s_i))\mathbbm{1}\left\lbrace y_t(s_i)<\hat{l}_t(s_i) \right\rbrace \\
 & +\frac{2}{\alpha}(y_t(s_i)-\hat{u}_t(s_i))\mathbbm{1}\left\lbrace y_t(s_i)>\hat{u}_t(s_i) \right\rbrace
\end{split}
\end{equation*}
where $\hat{l}_t(s_i)$ and $\hat{u}_t(s_i)$ are, respectively, the lower bound obtained by the $\frac{\alpha}{2}$ quantile, and the upper bound, obtained by the $1-\frac{\alpha}{2}$ quantile based on the predictive distribution. The indicator function is represented by $\mathbbm{1}$.

Smaller IS values indicate more efficient probabilistic forecasts.

\subsection{DIC}

Particularly useful in Bayesian model selection problems, the deviance information criterion \citep[DIC,][]{spiegelhalter2002bayesian} is a hierarchical modeling generalization of the Akaike information criterion \citep[AIC,][]{akaike1974new}. This criterion is negatively oriented implying that models with smaller DIC should be preferred to models with larger DIC. The DIC is given by:
$$ \text{DIC} = -2 \int\log\{p(\textbf{y}|\Theta)\}p(\Theta|\textbf{y})d\Theta + \text{P}_\text{D}, $$
where $\textbf{y}$ is a vector of observed values and $\Theta$ is the parameter vector. Thus, $p(\textbf{y}|\Theta)$ represents the likelihood function and $p(\Theta | \textbf{y})$, the posterior distribution. Defined as a Bayesian measure of model complexity, $\text{P}_\text{D}$ is given by:
$$ \text{P}_\text{D} = 2 \log\{p(\textbf{y}|\tilde{\Theta})\} -2 \int\log\{p(\textbf{y}|\Theta)\}p(\Theta|\textbf{y})d\Theta, $$
with $\tilde{\Theta}$ denoting the Bayes estimator. 

A Monte Carlo approximation of DIC is given by:
$$ \widehat{\text{DIC}} = -2 \frac{1}{M}\sum_{m = 1}^{M} \log\{p(\textbf{y}|\Theta^{(m)})\} + \hat{\text{P}}_\text{D}, $$
where $$\hat{\text{P}}_\text{D} = 2 \log\{p(\textbf{y}|\tilde{\Theta})\} -2 \frac{1}{M}\sum_{m = 1}^{M}\log\{p(\textbf{y}|\Theta^{(m)})\},$$
with ${\Theta^{(m)}}$ denoting the $m$-th MCMC sample of $\Theta$ from posterior distribution $p(\Theta|\textbf{y})$, $m = 1, ..., M$.

\subsection{LPML}

A component of the Bayes factor, the logarithm of the pseudo marginal likelihood \citep[LPML,][]{gelfand1996model} is given by:
$$ \text{LPML} = \frac{1}{n} \sum_{i=1}^{n}\log\{\text{CPO}_i\}, $$

\noindent
where $\text{CPO}_i$ represents the conditional predictive ordinate (CPO) for location $i$ and is given by:
\begin{equation*}
\begin{split}
\text{CPO}_i = & \quad p\big(\textbf{y}(s_i)| \textbf{y}(s_{-i})\big) \\
= & \quad \bigg(\int \frac{1}{p(\textbf{y}(s_i)|\Theta)}p(\Theta|\textbf{y})d\Theta\bigg)^{-1},
\end{split}
\end{equation*}
with $\textbf{y}(s_{i})=\big({y}_{1}(s_{i}), ..., {y}_{T}(s_{i})\big)'$ and $\textbf{y}(s_{-i})=\big(\textbf{y}(s_{1}),...,\textbf{y}(s_{i-1}), \textbf{y}(s_{i+1}), ..., \textbf{y}(s_{n})\big)'$, $i = 1,...,n.$ Note that the CPO$_i$ is based on the leave-one-out-cross-validation process and estimates the probability of $\textbf{y}(s_{i})$ given the observation of $\textbf{y}(s_{-i})$. In particular, a Monte Carlo approximation of CPO$_i$ is given by:
$$ \widehat{\text{CPO}}_i = \bigg(\frac{1}{M}\sum_{m = 1}^{M} \frac{1}{p(\textbf{y}(s_i)|\Theta^{(m)})}\bigg)^{-1}, $$
with ${\Theta^{(m)}}$ denoting the $m$-th MCMC sample of $\Theta$ from posterior distribution $p(\Theta|\textbf{y})$, $m = 1, ..., M$.

Finally, the Monte Carlo approximation of LPML is given by: 

$$ \widehat{\text{LPML}} = \frac{1}{n} \sum_{i=1}^{n}\log\{\widehat{\text{CPO}}_i\}. $$

The preferred model maximizes this criterion.

\newpage

\section{Supplementary results}
\label{append:sup}

\begin{figure}[ht]
 \centering
 \makebox{\includegraphics[scale=.6, angle=-90]{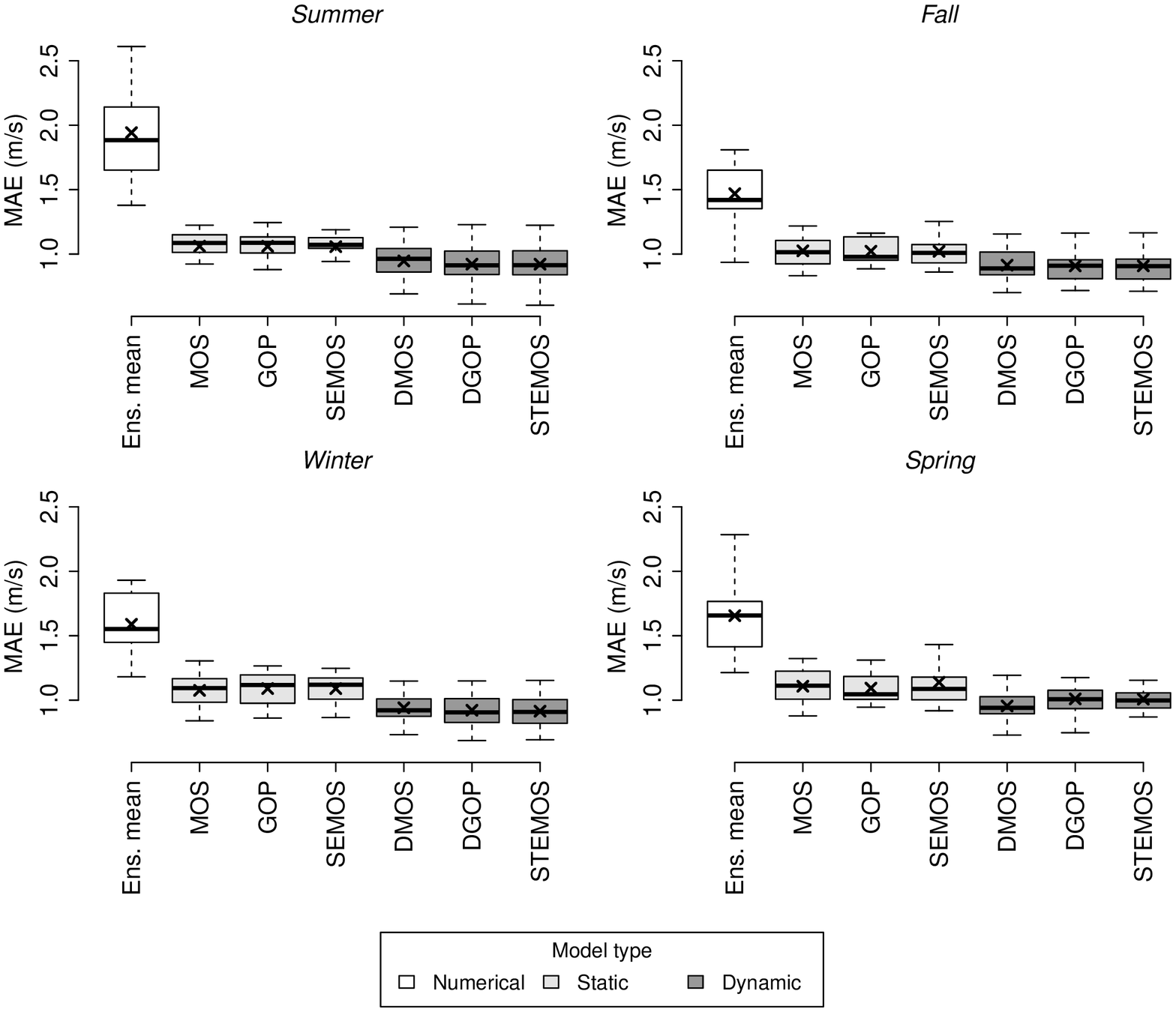}}
  \caption{Boxplots of average MAE of 24-hour deterministic forecasts for wind speed at 10-meter height during the seasons. The dense horizontal line and the $\mathbf{\times}$ symbol in each boxplot represent respectively, the median and mean values.}
\label{fig:MAE}
\end{figure}

\begin{figure}
 \centering
 \makebox{\includegraphics[scale=.6, angle=-90]{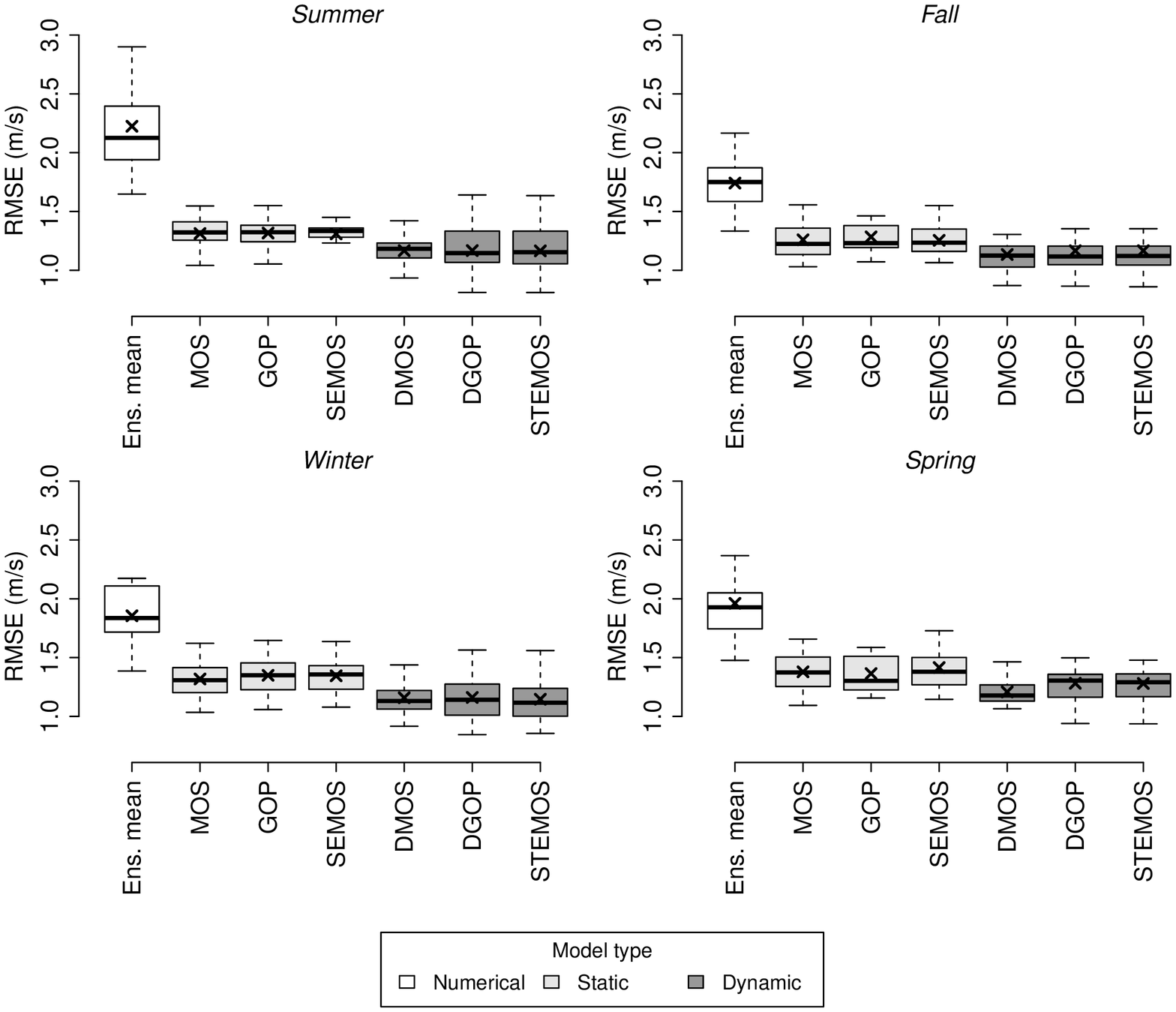}}
  \caption{Boxplots of average RMSE of 24-hour deterministic forecasts for wind speed at 10-meter height during the seasons. The dense horizontal line and the $\mathbf{\times}$ symbol in each boxplot represent respectively, the median and mean values.}
\label{fig:RMSE}
\end{figure}

\begin{figure}
 \centering
 \makebox{\includegraphics[scale=.6, angle=-90]{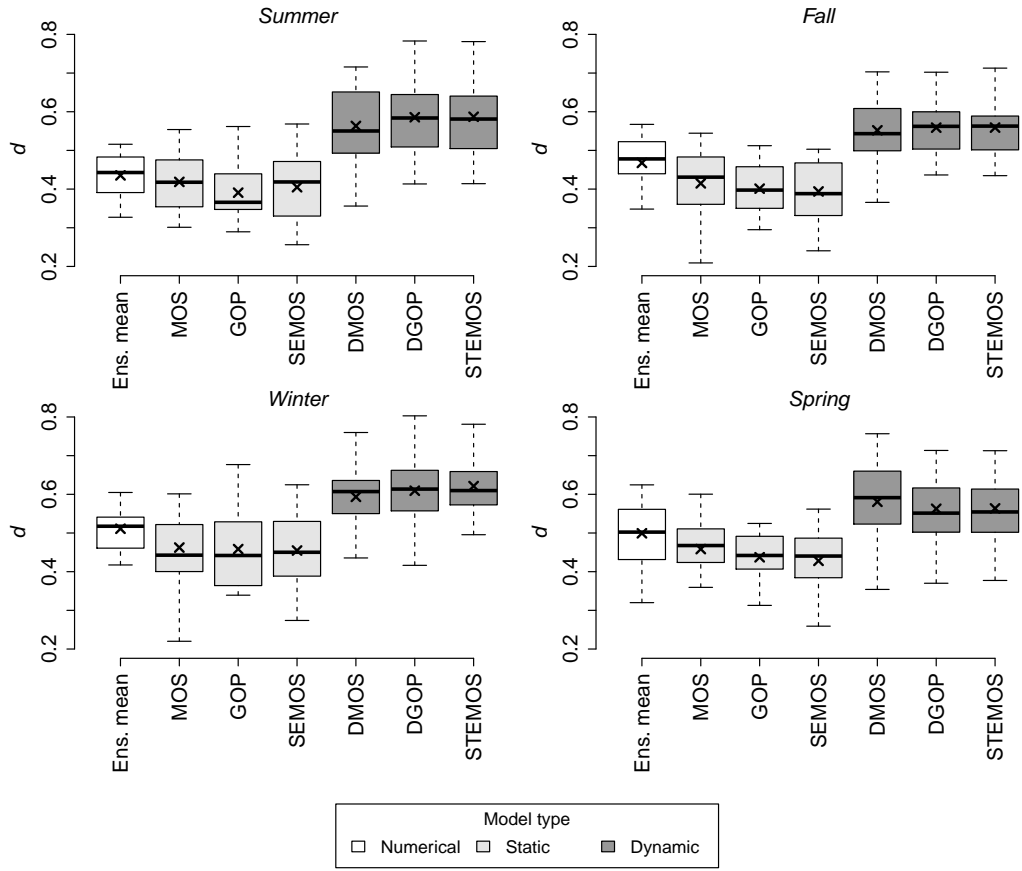}}
  \caption{Boxplots of average index of agreement of 24-hour deterministic forecasts for wind speed at 10-meter height during the seasons. Absolute agreement between predictions and actual values occurs when $d=1$. The dense horizontal line and the $\mathbf{\times}$ symbol in each boxplot represent respectively, the median and mean values.}
 \label{fig:d}
\end{figure}

\begin{figure}
 \centering
 \makebox{\includegraphics[scale=.7]{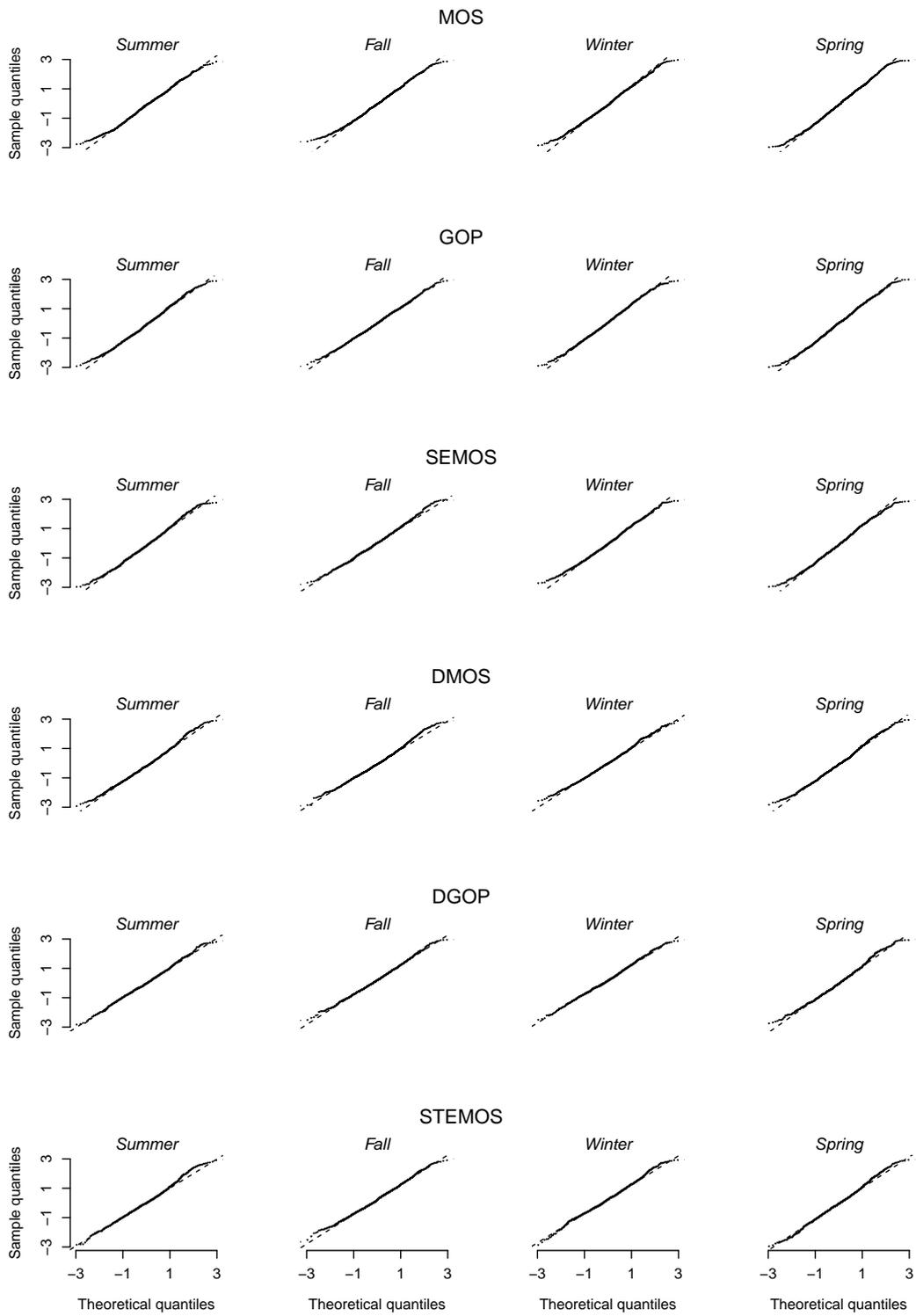}}
  \caption{Normal quantile-quantile plots for residuals of each fitted model during the seasons.}
 \label{fig:qqplot}
\end{figure}

\newpage

% --- References
\bibliographystyle{jasa}
\bibliography{main}

\end{document}